\documentclass[prb,twocolumn,showpacs]{revtex4}
\usepackage{amsmath,amssymb}
\usepackage{mathrsfs}
\usepackage{graphicx}

\begin{document}

\title{Kondo Effects in Carbon Nanotubes: From SU(4) to SU(2) symmetry}%

\author{Jong Soo Lim}%
\affiliation{Department of Physics, Seoul National University, Seoul
  151-747, Korea}%
\affiliation{Department of Physics, Korea University, Seoul 136-701,
  Korea}%
\author{Mahn-Soo Choi}%
\email{choims@korea.ac.kr}%
\affiliation{Department of Physics, Korea University, Seoul 136-701,
  Korea}%
\affiliation{Department de F\'isica, Universitat de les Illes Balears,
  E-07122 Palma de Mallorca, Spain }%
\author{M. Y. Choi}%
\affiliation{Department of Physics and Center for Theoretical Physics,
Seoul National University, Seoul 151-747, Korea}%
\affiliation{Korea Institute for Advanced Study, Seoul 130-722, Korea}%
\author{Rosa L\'opez}%
\affiliation{Department de F\'isica, Universitat de les Illes Balears,
  E-07122 Palma de Mallorca, Spain }%
\author{Ram\'on Aguado}%
\affiliation{Teor\'{\i}a de la Materia Condensada, Instituto de Ciencia
  de Materiales de Madrid (CSIC) Cantoblanco,28049 Madrid, Spain}%
\date{\today}

\begin{abstract}
We study the Kondo effect in a single-electron transistor device
realized in a single-wall carbon nanotube. The $K$-$K'$ double orbital
degeneracy of a nanotube, which originates from the peculiar
two-dimensional band structure of graphene, plays the role of a
pseudo-spin. Screening of this pseudo-spin, together with the real
spin, can result in an SU(4) Kondo effect at low temperatures. For
such an exotic Kondo effect to arise, it is crucial that this
orbital quantum number is conserved during tunneling.
Experimentally, this conservation is not obvious and some mixing
in the orbital channel may occur. Here we investigate in detail
the role of mixing and asymmetry in the tunneling coupling and
analyze how different Kondo effects, from the SU(4) symmetry to a
two-level SU(2) symmetry, emerge depending on the mixing and/or
asymmetry.  We use four different theoretical approaches to
address both the linear and non-linear conductance for different
values of the external magnetic field. Our results point out
clearly the experimental conditions to observe exclusively SU(4)
Kondo physics. Although we focus on nanotube quantum dots, our
results also apply to vertical quantum dots. We also mention that
a finite amount of orbital mixing corresponds, in the pseudospin
language, to having non-collinear leads with respect to the
orbital ''magnetization'' axis which defines the two pseudospin
orientations in the nanotube quantum dot. In this sense, some of
our results are also relevant to the problem of a Kondo quantum
dot coupled to non-collinear ferromagnetic leads.
\end{abstract}

\maketitle%

\let\eps=\epsilon%
\let\veps=\varepsilon%
\let\up=\uparrow%
\let\down=\downarrow%
\newcommand\eff{{\mathrm{eff}}}%
\newcommand\orb{{\mathrm{orb}}}%
\newcommand\sufour{{\mathrm{SU(4)}}}%
\newcommand\sutwo{{\mathrm{SU(2)}}}%
\newcommand\nm{\,\mathrm{nm}}%
\newcommand\K{\,\mathrm{K}}%
\newcommand\eV{\,\mathrm{eV}}%
\newcommand\mK{\,\mathrm{mK}}%
\newcommand\meV{\,\mathrm{meV}}%
\newcommand\half{\frac{1}{2}}%
\newcommand\bfs{\mathbf{s}}%
\newcommand\bfS{\mathbf{S}}%
\newcommand\bft{\mathbf{t}}%
\newcommand\bfT{\mathbf{T}}%
\newcommand\varN{\mathscr{N}}%
\newcommand\varS{\mathscr{S}}%
\newcommand\tilH{\widetilde{H}}%
\newcommand\tilV{\widetilde{V}}%
\newcommand\sign{\mathrm{sign}}%
\newcommand\avg[1]{\left\langle\textstyle#1\right\rangle}%
\newcommand\ket[1]{\left|\textstyle#1\right\rangle}
\newcommand\bra[1]{\left\langle\textstyle#1\right|}
\newcommand\varH{\,\mathscr{H}}

\section{Introduction}

The first observations of Kondo effect in semiconductor quantum
dots (QDs) \cite{Gold98,Cron98,Stutt98} have spurred a great deal
of experimental and theoretical activity during the last few
years. Since these experimental breakthroughs, remarkable
achievements have been reported, including the observation of the
unitary limit,\cite{Wiel00} the singlet-triplet Kondo effect,
\cite{Sasaki00a} Kondo effect in molecular conductors
\cite{molecule}, and the Kondo effect in QDs connected to
ferromagnetic \cite{Pasupathy04a} and superconducting reservoirs,
\cite{Buitelaar02a} just to mention a few.

Recently, Jarillo-Herrero {\it et al.} reported perhaps the most
sophisticated example, namely the observation of an orbital Kondo
effect in a carbon nanotube (CNT) quantum dot
(QD).\cite{Jarillo-Herrero05a} In these experiments it was shown
that the delocalized electrons of the reservoirs can screen both
the orbital pseudospin degrees of freedom in the CNT QD (the
$K$-K$'$ double orbital degeneracy of the two-dimensional band
structure of graphene) and the usual spin degrees of freedom,
resulting in an SU(4) Kondo effect at low temperatures. In a
recent letter, \cite{ChoiMS05z} we showed that quantum
fluctuations between the orbital and spin degrees of freedom may
indeed dominate transport at low temperatures and lead to this
highly symmetric SU(4) Kondo effect.
More recently, Sakano and Kawakami\cite{Sakano06a} have studied,
using the Bethe ansatz method at zero temperature and the
non-crossing approximation at finite temperatures, the more
general case where the quantum numbers of $N$ degenerate orbital
levels are conserved, and found new interesting features of the
SU($2N$)-symmetric Kondo effect.
Importantly, this is true provided that both the orbital
and spin indices are conserved during tunneling. This poses an
interesting question about the nature of the nanotube-lead contact
because, in principle, there is no special reason why the orbital
degrees of freedom in the CNT should be conserved during tunneling.

As mentioned, this orbital pseudospin originates from the peculiar
electronic structure of the nanotube (NT).
\cite{Jarillo-Herrero05a,Minot04a,Cao04a} The electronic states of
a NT form one-dimensional electron and hole sub-bands as a result
of the quantization of the electron wavenumber $k_{\perp}$
perpendicular to the NT axis, which arises when graphene is
wrapped into a cylinder to create a NT. By symmetry, for a given
sub-band at $k_{\perp}=k_0$ there is a second degenerate sub-band
at $k_{\perp}=-k_0$. Semiclassically, this orbital degeneracy
corresponds to the clockwise ($\circlearrowright$) or
counterclockwise ($\circlearrowleft$) symmetry of the wrapping
modes. A plausible explanation of why this degree of freedom is
preserved during tunneling could be that the QD is likely coupled
to NT electrodes (the metal electrodes are deposited on top of the
NT so maybe the electrons tunneling out of the QD enter the NT
section underneath the contacts) but this issue clearly deserves a
thorough microscopic analysis about
the nature of the contacts. 
The conservation of the orbital quantum number seems more likely
in the vertical quantum dots (VQD),\cite{Sasaki04a} where the
orbital quantum number is the magnetic quantum number of the
angular momentum.

Here, we take a different route and, assuming some degree of mixing in
the orbital channel, ask ourselves about the robustness of the SU(4)
Kondo effect against asymmetry in the couplings and/or mixing.

The rest of the paper is organized as follows: In
Section~\ref{sec:model} we introduce the relevant model
Hamiltonian and classify different schemes of the lead-dot
coupling. These different coupling schemes result in different
symmetries and hence affect significantly the underlying Kondo
physics. These effects are analyzed in the subsequent sections.
Section~\ref{sec:RG} presents the analysis with two
renormalization group (RG) approaches. In Sec.~\ref{sec:SBT} two
slave-boson approaches complement the previous results. Finally,
Sec.~\ref{sec:conclusion} concludes the paper.

\section{Model}
\label{sec:model}

\subsection{Nearly Degenerate Localized Orbitals}
We consider a QD with two (nearly) degenerate localized orbitals
which is coupled to reservoirs. As we mentioned before, we have in
mind the experimental setup of Ref.~\onlinecite{Jarillo-Herrero05a} where a highly symmetric
Kondo effect was demonstrated in a CNT QD. However, our description
could well apply to vertical quantum dot (VQD),\cite{Sasaki04a}
where the orbitals correspond to two degenerate Fock-Darwin states
with different values of the angular momentum quantum number.
Hereafter we denote this orbital quantum number by $m=1,2$.
The dot is then described by the Hamiltonian
\begin{subequations}
\label{su4vs2::eq:H1}
\begin{multline}
\label{su4vs2::eq:HD}
H_D = \sum_{m=1,2}\sum_{\sigma=\up,\down}
\eps_{m\sigma} d_{m\sigma}^\dag d_{m\sigma} \\\mbox{}%
+ \sum_{(m,\sigma)\neq(m',\sigma')} U_{mm'}n_{m\sigma}n_{m'\sigma'} \,,
\end{multline}
where $\eps_{m\sigma}$ is the single-particle energy level of the
localized state with orbital $m$ and spin $\sigma$, $d_{m\sigma}^\dag$
($d_{m\sigma}$) the fermion creation (annihilation) operator of the
state,
\begin{math}
n_{m\sigma} = d_{m\sigma}^\dag d_{m\sigma}
\end{math}
the occupation number operator, $U_{mm}$ ($m=1,2$) the intra-orbital Coulomb
interaction, and $U_{12}$ the inter-orbital Coulomb interaction.
The effect of the external magnetic field parallel to the symmetry
axis of the system is to lift the orbital and spin degeneracy
of the single-particle energy levels.  We will denote them by
$\Delta_\orb$ and $\Delta_Z$, respectively, so that the
single-particle energy levels $\eps_{m\sigma}$ have the form
\begin{equation}
\label{su4vs2::eq:1-2} \eps_{m\sigma} = \eps_0 +
\Delta_\orb(\delta_{m,1}-\delta_{m,2}) +
(\Delta_Z/2)(\delta_{\sigma,\up}-\delta_{\sigma,\down})\,.
\end{equation}
The precise values of the Coulomb interactions $U_{mm'}$ depend on
the details of the system, but should be of the order of the
charging energy $E_C= e^2/2C$ with $C$ being the total capacitance
of the dot.  In this work we focus on the regime where the system
of the localized levels is occupied by a single electron
($\sum_{m\sigma}\avg{n_{m\sigma}}\approx 1$, quarter filling
\cite{half-filling}) and the Coulomb interaction energy
($U_{mm'}\sim E_C$) is much bigger than other energy scales.  In
this regime the Hamiltonian in Eq.~(\ref{su4vs2::eq:HD}) suffices
to describe all relevant physics of our concern.

\subsection{Coupling Schemes}

Kondo physics arises as a result of the interplay between strong
correlations in the dot and coupling of the localized electrons
with the itinerant ones in conduction bands.  Naturally, different
Kondo effects are observed depending on the way the dot is coupled
to the electrodes and whether or not the orbital quantum number
$m$ is conserved. Nevertheless, it turns out highly non-trivial
\emph{experimentally} to distinguish those different Kondo
effects.  In subsequent sections we will consider different
coupling schemes between the dot and the electrodes, show how
different physics emerges, and propose how to distinguish them
unambiguously in experiments.

\begin{figure}
\centering
\includegraphics*[width=8cm]{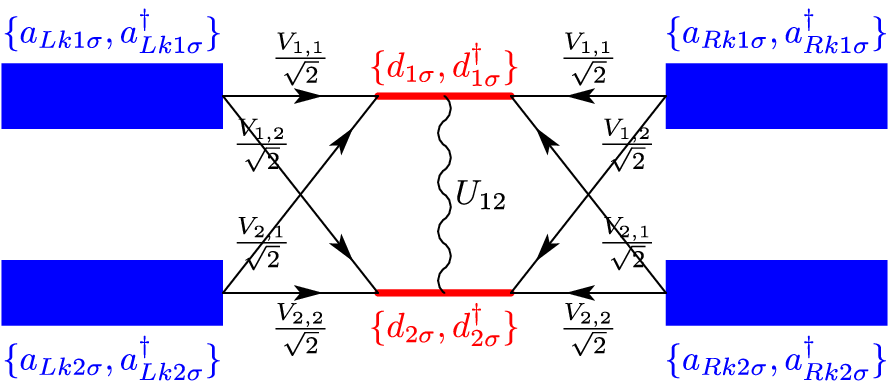}
\includegraphics*[width=7cm]{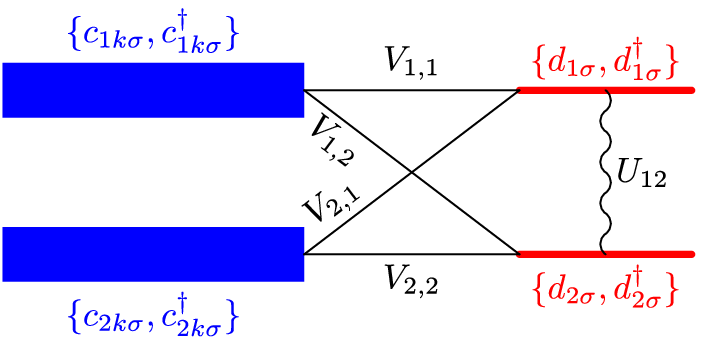}
\caption{(color online) Schematic of a representative mesoscopic system
  in question.  In (a) Each of the two leads, $L$ and $R$, has two
  conduction bands (or ``modes''), $1$ and $2$.  The model with two
  leads in (a) is equivalent in equilibrium to the model in (b) with
  only one lead.  The operators $c_{k\mu\sigma}$ ($\mu=1,2$ and
  $\sigma=\up,\down$) are related to $a_{Lk\mu\sigma}$ and
  $a_{Rk\mu\sigma}$ by the canonical transformation in
  Eq.~(\ref{su4vs2::eq:CT1}).  The wiggly lines indicate the
  inter-orbital Coulomb interaction $U_{12}$ whereas the intra-orbital
  interaction $U_{mm}$ ($m=1,2$) is not shown.}
\label{su4vs2::fig:model-1,2}
\end{figure}

The two leads $\alpha=L$ and $R$ are treated as non-interacting gases of
fermions:
\begin{equation}
\label{su4vs2::eq:HC1}
H_\alpha =
\sum_k\sum_{\mu=1,2}\sum_\sigma\eps_{\alpha k \mu}\,
a_{\alpha k\mu\sigma}^\dag a_{\alpha k\mu\sigma} \,,
\end{equation}
where $\mu$ denotes channels in the leads.  Without loss of
generality, we assume that there are two distinguished (groups of)
channels $\mu=1$ and $2$ in each lead.  When the leads bears the
same symmetry as the dot, this channel quantum number $\mu$ in the
leads is identical to the orbital quantum number $m$ in the dot
and will be preserved over the tunneling of electrons from the dot
to leads and vice versa; see Fig.~\ref{su4vs2::fig:model-1,2}(a).
Otherwise, the orbital channels become mixed.
The most general situation is described by the tunneling Hamiltonian
\begin{equation}
\label{su4vs2::eq:HT1}
H_T =
\sum_{\alpha k\mu\sigma}
\left(V_{\alpha k\mu m\sigma}
  a_{\alpha k\mu\sigma}^\dag d_{m\sigma}
  + h.c.\right)
\end{equation}
\end{subequations}
and the total Hamiltonian is thus given by
\begin{math}
H = H_L + H_R + H_T + H_D
\end{math}\,.

For the sake of simplicity, we assume identical electrodes
($\eps_{Lk\mu}=\eps_{Rk\mu}$) together with symmetric tunneling
junctions ($V_{Lk\mu m\sigma}=V_{Rk\mu m\sigma}$), and ignore
their $k$- and $\sigma$-dependence of the tunneling amplitudes. In
this way, we consider a simplified model with $V_{\alpha k\mu
m\sigma}=V_{\mu,m}/\sqrt{2}$ and define the widths
\begin{equation} \Gamma_m \equiv \Gamma_{mm}
\,,\quad \Gamma_{mm'} = \pi\rho_0 V_m^* V_{m'}^{} \,
\end{equation}
with $V_m\equiv V_{m,m}$, where $\rho_0 $ is the density of states
(DOS) in the reservoirs. Then \emph{in equilibrium} the
Hamiltonian $H$ in Eq.~(\ref{su4vs2::eq:H1}) is equivalent to
$H=H_C+H_T+H_D$ with
\begin{subequations}
\label{su4vs2::eq:H2}
\begin{align}
\label{su4vs2::eq:HC2}
H_C & =\sum_{k\mu\sigma}\veps_{k\mu}c_{k\mu\sigma}^\dag c_{k\mu\sigma} \,,\\
\label{su4vs2::eq:HT2}
H_T & = \sum_{k\mu m \sigma}
\left(V_{\mu , m}c_{k\mu\sigma}^\dag d_{m\sigma} + h.c.\right) \,,
\end{align}
\end{subequations}
where we have performed the canonical transformation
\begin{equation}
\label{su4vs2::eq:CT1}
\begin{split}
c_{k\mu\sigma} & = \frac{a_{Lk\mu\sigma} + a_{Rk\mu\sigma}}{\sqrt{2}} \,,\\
b_{k\mu\sigma} & = \frac{a_{Lk\mu\sigma} -
a_{Rk\mu\sigma}}{\sqrt{2}} \,,
\end{split}
\end{equation}
and discarded the decoupled term
\begin{math}
\eps_{k\mu} b_{k\mu\sigma}^\dag b_{k\mu\sigma}
\end{math}.

In the following sections we investigate the physics described by
the Hamiltonian in Eq.~(\ref{su4vs2::eq:H2}) and, in particular,
clarify the role of index conservation in the symmetry of the
underlying Kondo regime at low temperatures. In order to carry out
this analysis, we use four different approaches: the scaling
theory (perturbative RG approach), the numerical renormalization
group (NRG) method, the slave-boson mean-field theory (SBMFT), and
the non-crossing approximation (NCA).

\section{Renormalization Group Approaches}
\label{sec:RG}

The renormalization group (RG) theory provides a convenient and
powerful method to study low-energy properties of strongly
correlated electron systems.  Here we take two RG approaches, the
scaling theory\cite{Anderson70a,Haldane78a,Haldane78b} and the NRG
method.~\cite{Wilson75a,Krishna-murthy80a,Costi94a,Hofstetter00a}
While the scaling theory is useful for qualitative understanding
of the model, a more precise quantitative analysis requires the
use of more sophisticated methods like the NRG method. This method
is known to be one of the most accurate and powerful theoretical
tools to study quantum impurity problems (see Appendix A).

\subsection{SU(4) Kondo Effect}

\begin{figure}
\centering
\includegraphics*{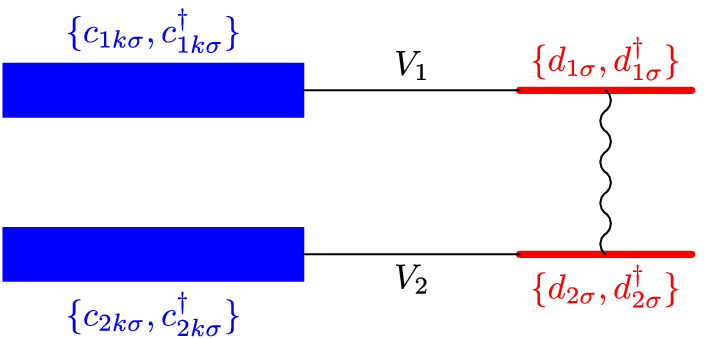}
\caption{(color online) Schematics of the SU(4)-symmetric Anderson
model.} \label{su4vs2::fig:model-4}
\end{figure}

We now turn to the case where tunneling processes conserve the
orbital quantum number; see Fig.~\ref{su4vs2::fig:model-4}. In
this case, the Hamiltonian reads
\begin{multline}
\label{su4vs2::eq:Hsu4-1} H =
\sum_{\alpha=L,R}\sum_{m=1,2}\sum_{k\sigma} \veps_{\alpha
k}a_{\alpha km\sigma}^\dag a_{\alpha km\sigma}
\\\mbox{}%
+ \sum_{\alpha km\sigma} V_{m} \left(a_{\alpha km\sigma}^\dag
d_{m\sigma}
  + d_{m\sigma}^\dag a_{\alpha km\sigma}\right)
+ H_D
\end{multline}
or [see Eqs.~(\ref{su4vs2::eq:HC2}) and (\ref{su4vs2::eq:HT2})]
\begin{multline}
\label{su4vs2::eq:Hsu4-2} H =
\sum_{km\sigma}\eps_kc_{k\sigma}^\dag c_{km\sigma}
\\\mbox{}%
+ \sum_{km\sigma}V_m\left(c_{km\sigma}^\dag d_{m\sigma} +
  d_{m\sigma}^\dag c_{km\sigma}\right) + H_D \,.
\end{multline}

From the RG point of view, starting initially with nearly
degenerate levels, all the localized levels are relevant for the
spin and orbital fluctuations, and, as we shall see below,
contribute to the Kondo effect.  To investigate the low-energy
properties of the orbital and spin fluctuations of the model, we
perform the Schrieffer-Wolf (SW) transformation and obtain an effective
Kondo-type Hamiltonian:
\begin{multline}
\label{su4vs2::eq:Hsu4-3} H =
\sum_{km\sigma}\eps_kc_{km\sigma}^\dag c_{km\sigma} +
H_\eff^\sufour - \Delta_Z S^z
- 2\Delta_\orb T^z  \\\mbox{}%
- \frac{\left(\sqrt{J_1}-\sqrt{J_2}\right)^2}{2}
(1+4\bfs\cdot\bfS)(t^xT^x+t^yT^y) \\\mbox{} +
(J_1-J_2)(\bfs\cdot\bfS)(t^z+T^z) \,,
\end{multline}
where
\begin{equation}
\label{su4vs2::eq:su4eff-1}
H_\eff^\sufour =
\frac{J_1+J_2}{2}\left[
  \bfs\cdot\bfS + \bft\cdot\bfT + 4(\bfs\cdot\bfS)(\bft\cdot\bfT)
\right] \,
\end{equation}
and the exchange coupling constants $J_m$ ($m=1,2$) are given by
\begin{equation}
\label{su4vs2::eq:J:1} J_m =
V_m^2\left(\frac{1}{E_+}+\frac{1}{E_-}\right) \,.
\end{equation}
We note that the Kondo-type effective Hamiltonian in
Eq.~(\ref{su4vs2::eq:Hsu4-3}) reduces to the SU(4)-symmetric
Kondo model when $V_1=V_2$ and $\eps_{1\sigma}=\eps_{2\sigma}$.
In this case, orbitals play exactly the same role as spins;
the former are not distinguished from the latter.

Under the RG transformation reducing subsequently the conduction
band width $D$ by $\delta{D}$, the Kondo-type effective
Hamiltonian evolves into a generic form:
\begin{multline}
\label{su4vs2::eq:3-7}
H_\eff = H_\mathrm{leads} -\Delta_Z S^z - 2\Delta_\orb T^z \\\mbox{}%
+ 2J_1(\bfs\cdot\bfS)\left(\half+t^z\right)\left(\half+T^z\right) \\\mbox{}%
+ 2J_2(\bfs\cdot\bfS)\left(\half-t^z\right)\left(\half-T^z\right) \\\mbox{}%
+ \half\left[J_4 + 4J_3(\bfs\cdot\bfS)\right](t^+T^- + t^-T^+) \\\mbox{}%
+ J_5t^zT^z \,.
\end{multline}
The level splitting $\Delta_\orb$ and $\Delta_Z$ remain constant
under the RG transformation:
\begin{equation}
\label{su4vs2::eq:3-8} \frac{d\Delta_Z}{d\ln D} =
\frac{d\Delta_\orb}{d\ln D} = 0 \,.
\end{equation}
The exchange coupling constants $J_i$ ($i=1,\cdots,5$) are
initially given by Eq.~(\ref{su4vs2::eq:J:1}) and
\begin{equation}
\label{su4vs2::eq:J:2} J_3 = J_4 = \sqrt{J_1J_2} \,,\quad J_5 =
\frac{1}{2} (J_1+J_2) \,,
\end{equation}
which, under the RG transformation, scale as
\begin{subequations}
\label{su4vs2::eq:rg:1}
\begin{align}
\label{su4vs2::eq:rg:1a}
\frac{dJ_1}{\rho_0 d\ln D} &= -2J_1^2-J_3(J_3+J_4) \,,\\
\label{su4vs2::eq:rg:1b}
\frac{dJ_2}{\rho_0 d\ln D} &= -2J_2^2-J_3(J_3+J_4) \,,\\
\label{su4vs2::eq:rg:1c}
\frac{dJ_3}{\rho_0 d\ln D} &= -J_3(J_1+J_2+J_5)-\frac{1}{2} J_4(J_1+J_2) \,,\\
\label{su4vs2::eq:rg:1d}
\frac{dJ_4}{\rho_0 d\ln D} &= -\frac{3}{2} J_3(J_1+J_2)-J_4J_5 \,,\\
\label{su4vs2::eq:rg:1e} \frac{dJ_5}{\rho_0 d\ln D} &=
-3J_3^2-J_4^2 \,
\end{align}
\end{subequations}
for $D\gg\Delta_\orb\geq\Delta_Z$.  For $D\ll\Delta_\orb$, it is
clear from Eq.~(\ref{su4vs2::eq:3-7}) that the orbital
fluctuations are frozen and only $J_1$ is relevant, which scales
as
\begin{equation}
\label{sejo} \frac{dJ_1}{\rho_0 d\ln D} = -2J_1^2.
\end{equation}
It implies that we recover the single-level Anderson model for
$D\ll\Delta_\orb$.   Therefore in the remainder of this section,
we will focus on the case $D\gg\Delta_\orb$.

It is convenient to define the reduced variables $j_i\equiv
J_i/J_1$ ($i=2,\cdots,5$) and rewrite the RG
equations~(\ref{su4vs2::eq:rg:1}) as
\begin{subequations}
\label{su4vs2::eq:rg:2}
\begin{align}
\frac{dj_2}{dx} &= -j_2 +
\frac{2j_2^2+j_3(j_3+j_4)}{2+j_3(j_3+j_4)},
\label{seq2} \\
\frac{dj_3}{dx} &= -j_3 + \frac{j_3(1+j_2+j_5)+(j_4 /2)
(1+j_2)}{2+j_3(j_3+j_4)},
\label{seq3} \\
\frac{dj_4}{dx} &= -j_4 + \frac{(3j_3 /2)
(1+j_2)+j_4j_5}{2+j_3(j_3+j_4)},
\label{seq4} \\
\frac{dj_5}{dx} &= -j_5 + \frac{3j_3^2+j_4^2}{2+j_3(j_3+j_4)} \,
\label{seq5}
\end{align}
\end{subequations}
with $x=\ln(\rho_0J_1)$, while $J_1$ obeys the scaling equation
\begin{equation}
\label{sejs} \frac{1}{(\rho_0J_1)^2}\frac{d(\rho_0J_1)}{d\ln D} =
-2-j_3(j_3+j_4) \,.
\end{equation}
The RG equations~(\ref{su4vs2::eq:rg:2}) have two fixed points:
one describing the SU(4) Kondo physics
\begin{equation}
\label{su4vs2::eq:fixedpoint:1a} j_2=j_3=j_4=j_5=1
\end{equation}
and the other describing the usual SU(2) Kondo physics
\begin{equation}
\label{su4vs2::eq:fixedpoint:1b} j_2=j_3=j_4=j_5=0 \,,
\end{equation}
both with $J_1=\infty$ as indicated in
Fig.~\ref{su4vs2::fig:rgflow-1,2}.  Linearizing the RG
equations~(\ref{su4vs2::eq:rg:2}) around the fixed points, one can
easily show that both the SU(2) and SU(4) Kondo fixed points are
stable (there is one marginal parameter at the SU(4)
fixed point). However, as indicated as a dashed semicircle in
Fig.~\ref{su4vs2::fig:rgflow-1,2}(b), the radius of convergence is
finite while the fixed point itself is located at infinity.  This
implies that in priciple, the SU(4) Kondo fixed point cannot be
reached for arbitrarily small values of $1-\Gamma_2/\Gamma_1$.
However, as illustrated in Fig.~\ref{su4vs2::fig:rgflow-1,2}(a),
in the region of physical interest for sufficiently small values
of $1-\Gamma_2/\Gamma_1$, the scaling behavior is essentially
governed by the SU(4) Kondo fixed point (see also
Fig.~\ref{su4vs2::fig:nrg-4}).
More importantly, for sufficiently small values of
$1-\Gamma_2/\Gamma_1$, the SU(2) fixed point governs the physics only at
extremely low energies.  This suggests that the SU(4) Kondo signature
can be observed exclusively at relatively higher energy scales (of the order
of the Kondo temperature), as in the experiment reported recently.~\cite{Jarillo-Herrero05a}

\begin{figure}
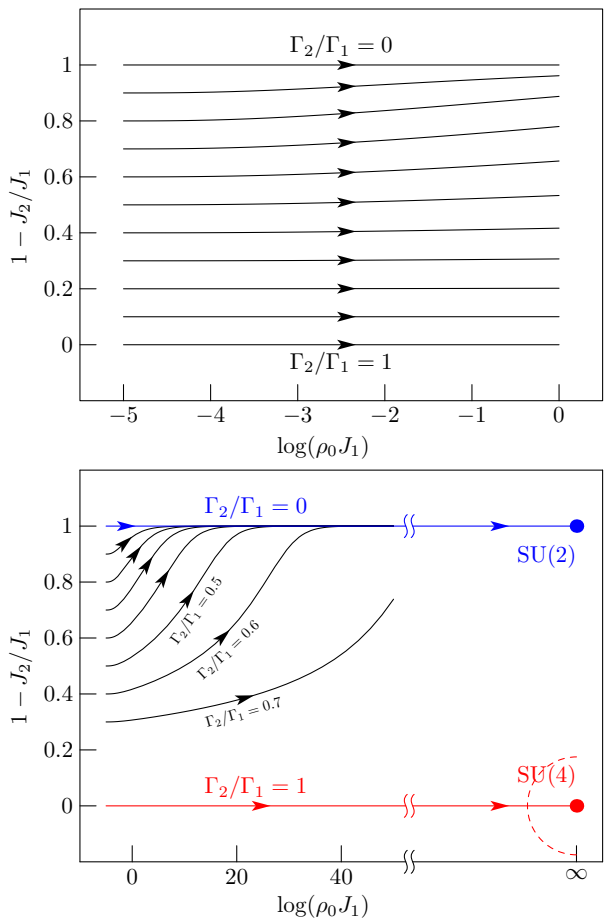

\centering
\includegraphics*[width=8cm]{fig-03a}
\includegraphics*[width=8cm]{fig-03b}
\caption{(color online) RG flows for different values of
  $\Gamma_2/\Gamma_1$ with $\Gamma_1$ fixed.}
\label{su4vs2::fig:rgflow-1,2}
\end{figure}

\begin{figure}
\centering
\includegraphics*[width=8cm]{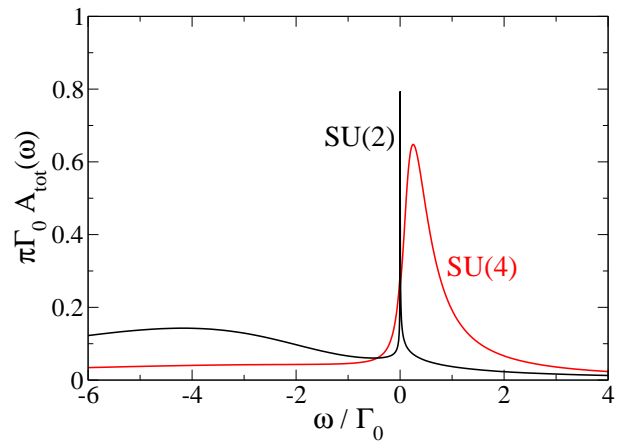}
\caption{Comparison of the SU(2) and SU(4) Kondo model.}
\label{su4vs2::fig:nrg-2}
\end{figure}

\begin{figure}
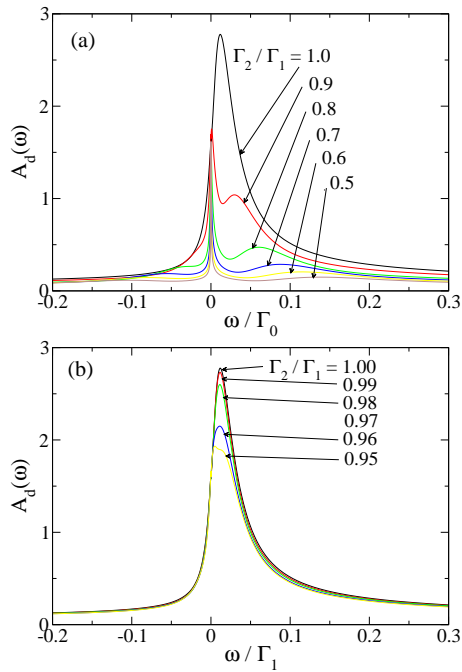

\centering
\includegraphics*[width=6cm]{fig-05a}
\includegraphics*[width=6cm]{fig-05b}
\caption{(color online) NRG results of the total spectral density
  $A_d(E)$ for different values of coupling asymmetry
  $\Gamma_2/\Gamma_1$.  The parameter values are: $\eps_{0}=-0.8D$,
  $\Gamma_1=0.1D$, $U_{mm'}=8D$, and $\Delta_\orb=\Delta_Z=0$.}
\label{su4vs2::fig:nrg-4}
\end{figure}

At $B_{\parallel}=0$ and $\Gamma_1=\Gamma_2\equiv\Gamma_0$, the RG
equations~(\ref{su4vs2::eq:rg:1}) reduce to a single equation
\begin{equation}
\label{su4vs2::eq:rg3} \frac{dJ_1}{\rho_0 d\ln D} = -4J_1^2.
\end{equation}
Comparing this with the corresponding one in Eq.~(\ref{sejo})
for the usual single-level Anderson model, we note that the Kondo
temperature is enhanced exponentially:
\begin{equation}
\label{su4vs2::eq:TKsu4} T_K^{SU(4)} \sim \exp(-1/4\rho_0J_1) \,
\end{equation}
with respect to the SU(2) Kondo temperature
\begin{equation}
\label{su4vs2::eq:TKsu2} T_K^{SU(2)} \sim \exp(-1/2\rho_0J_1) \,.
\end{equation}
The perturbative RG analysis discussed above, whose validity is
guaranteed only for $\rho_0J_i\ll 1$, turns out to be
qualitatively correct in a wide region of the parameter space and
provides a clear interpretation of the model.  To confirm the
perturbative RG analysis and make quantitative analysis, we use
the NRG method (described in Appendix A), the results of which
are summarized in Figs.~\ref{su4vs2::fig:nrg-2}--\ref{su4vs2::fig:nrg-4}.
There the \emph{total} spectral density
\begin{equation}
\label{su4vs2::eq:2-23} A_d(E) =
\sum_\sigma\sum_{mm'}\pi\Gamma_{mm'}A_{m'm;\sigma}(E) \,,
\end{equation}
which provides direct information on the linear conductance~\cite{Meir92a},
is plotted.

The spectral density shows a peak near the Fermi energy,
corresponding to the formation of the SU(4) Kondo state (see
Fig.~\ref{su4vs2::fig:nrg-2}). The peak width, which is much
broader than that for the SU(2) Kondo model (represented by the dotted line),
demonstrates the exponential enhancement of the Kondo temperature
mentioned above. Another remarkable effect is that the SU(4) Kondo
peak shifts away from $\omega=E_F=0$ and is pinned at
$\omega\approx T_K^\mathrm{SU(4)}$.  This can be understood from
the Friedel sum rule,~\cite{Langreth66a} which in this case gives
$\delta=\pi/4$ for the scattering phase shift at $E_F$.
Accordingly, the linear conductance at zero temperature is given
by $\mathcal{G}_0 = 4(e^2/h)\sin^2\delta = 2e^2/h$. It is
interesting to recall that the Friedel sum rule gives the same
linear conductance also for the two-level SU(2) Kondo model. Thus,
neither the enhancement of the Kondo temperature nor the linear
conductance can distinguish between the SU(4) and the two-level
SU(2) Kondo effects.  This can be achieved only by studying the
influence of a parallel magnetic field in the nonlinear
conductance, as shown in Ref. \onlinecite{Jarillo-Herrero05a}

\subsection{Effects of Mixing of Orbital Quantum Numbers}

\begin{figure}
\centering
\includegraphics*{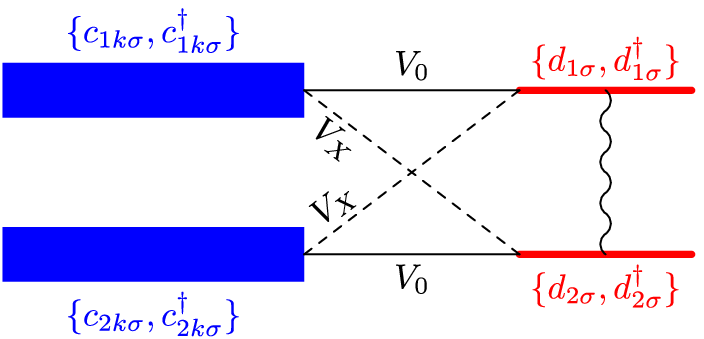}
\caption{(color online) Model with finite mixing between orbital
  quantum numbers.}
\label{su4vs2::fig:model-5}
\end{figure}

To examine the stability of the SU(4) Kondo phenomena against
orbital mixing, we consider the model (see
Fig.~\ref{su4vs2::fig:model-5}):
\begin{multline}
\label{su4vs2::eq:H4-1} H = \sum_{k\sigma}\veps_kc_{k\sigma}^\dag
c_{k\sigma} + \sum_{km\sigma}V_0\left(c_{km\sigma}^\dag
d_{m\sigma} +
  d_{m\sigma}^\dag c_{km\sigma}\right) \\\mbox{}%
+ \sum_{km\sigma}V_X\left(c_{k\bar{m}\sigma}^\dag d_{m\sigma} +
  d_{m\sigma}^\dag c_{k\bar{m}\sigma}\right)
+ H_D \,,
\end{multline}
where the indices imply $\bar{1}=2$ and
$\bar{2}=1$ and $V_0\equiv V_{1,1}=V_{2,2}$ and $V_X\equiv
V_{1,2}=V_{2,1}$.

If we rewrite the Hamiltonian in the form
\begin{eqnarray}
\label{su4vs2::eq:H4-1b} H &=&
\sum_{k\sigma}\veps_kc_{k\sigma}^\dag c_{k\sigma} +
\sum_{km\sigma}\{V_0c_{km\sigma}^\dag+V_Xc_{k\bar{m}\sigma}^\dag\}
d_{m\sigma} \nonumber\\
&+&  d_{m\sigma}^\dag \{V_0c_{km\sigma}+ V_Xc_{k\bar{m}\sigma}\} +
H_D \,,
\end{eqnarray}
it now becomes clear that, in the pseudospin language, a finite
amount of orbital mixing corresponds to having non-collinear leads
with respect to the orbital ``magnetization'' axis which defines
the pseudospin orientations $m=1$ and $m=2$ in the dot. In other
words, each confined electron (with defined pseudospin) couples to
a linear combination of pseudospins and, as a result, becomes
rotated in the pseudospin space by the angle defined by
$\tan\phi=V_X/V_0$. Note that for the maximal mixing $V_X=V_0$,
the tunneling electrons lose completely information about their
pseudospin orientation. In this limit, one recovers the spin Kondo
physics [with SU(2) symmetry] of a two-level Anderson model [see the next
subsection and Eq.~(\ref{su4vs2::eq:Hsu2-2})].  For zero mixing
($V_X=0$), the model reduces to the SU(4)-symmetric model of
Eq.~(\ref{su4vs2::eq:Hsu4-2}) (with tunneling amplitudes which do not
depend on the orbital index).

After the RG transformation of the Anderson-type model in
Eq.~(\ref{su4vs2::eq:H4-1}) until the single-particle energy
levels are comparable with the conduction band width (when the
charge fluctuations are suppressed), the SW transformation gives
\begin{multline}
\label{su4vs2::eq:Mixing1} H_\eff = \left(1 -
\frac{J_X}{J_0}\right)H_\eff^\sufour
+ \frac{J_X}{J_0} H_\eff^\sutwo \\\mbox{}%
+ J_0\sqrt{\frac{J_X}{J_0}} \left(1 -
\sqrt{\frac{J_X}{J_0}}\right)(1+4\bfs\cdot\bfS)(t^x + T^x)
\\\mbox{}%
+ 2J_X(t^xT^x)  \,,
\end{multline}
where
\begin{equation}
\label{su4vs2::eq:su4eff} H_\eff^\sufour = J_0\left[
  \bfs\cdot\bfS + \bft\cdot\bfT + 4(\bfs\cdot\bfS)(\bft\cdot\bfT)
\right]
\end{equation}
corresponds to the SU(4) Kondo model and
\begin{equation}
\label{su4vs2::eq:su2eff} H_\eff^\sutwo =
2J_0\bfs\cdot\bfS(1+2t^x)(1+2T^x) + J_0(t^x+T^x) \,.
\end{equation}
the SU(2) Kondo model.  The exchange coupling constants $J_0$ and
$J_X$, respectively, are given by
\begin{equation}
\label{su4vs2::eq:J0} J_0 = |V_0|^2\left(\frac{1}{E_+} +
\frac{1}{E_-}\right) \,,\quad J_X = |V_X|^2\left(\frac{1}{E_+} +
\frac{1}{E_-}\right) \,.
\end{equation}
One can already grasp an idea about the effects of the mixing
$J_X$ (i.e., $\Gamma_X$) of the orbital quantum numbers by
considering the two limiting cases, $J_X=0$ (no mixing) and
$J_X=J_0$ (maximal mixing), of the effective Hamiltonian
(\ref{su4vs2::eq:Mixing1}).  In the case of no mixing ($J_X=0$),
the effective Hamiltonian (\ref{su4vs2::eq:Mixing1}) reduces to
the SU(4)-symmetric Kondo model in Eq.~(\ref{su4vs2::eq:su4eff}),
which has been already discussed in the previous section: The Kondo
temperature is given by $T_K\sim D\exp(-1/4J_0)$.  When the mixing is
maximal ($J_X=J_0$), on the other hand, the effective Hamiltonian
becomes $H_\eff^\sutwo$ in Eq.~(\ref{su4vs2::eq:su2eff}).

\begin{figure}
\centering
\includegraphics*[width=8cm]{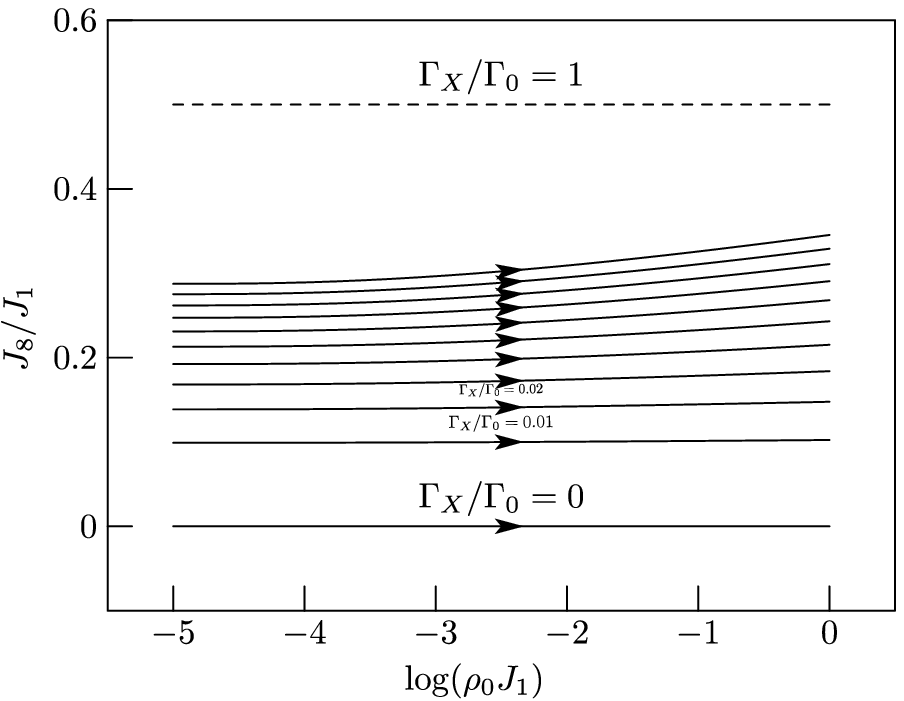}
\includegraphics*[width=8cm]{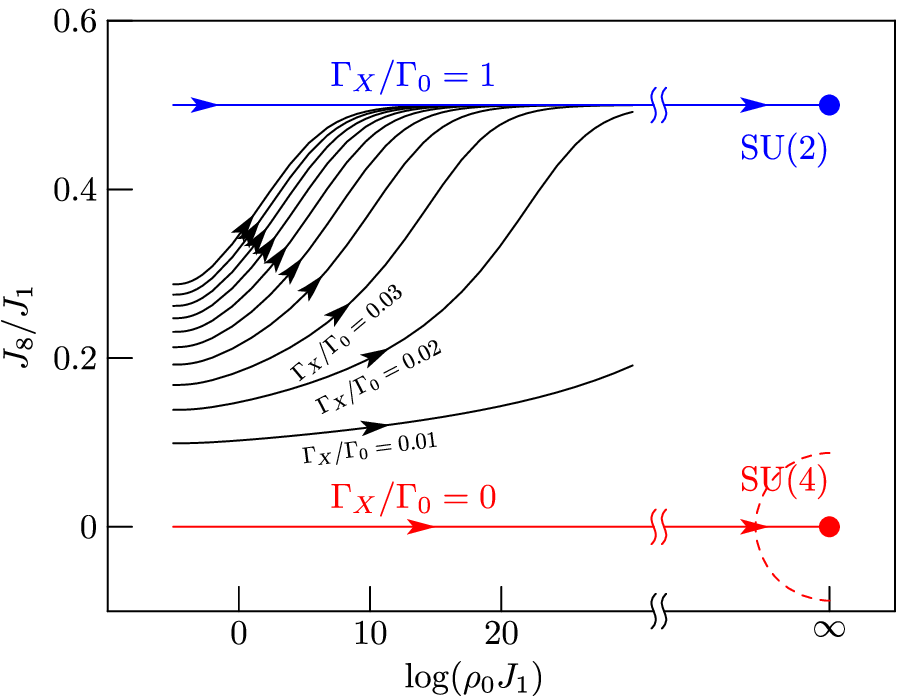}
\caption{(color online) RG flows in case that there is a finite amount of
  mixing of the orbital quantum numbers.}
\label{su4vs2::fig:rgflow-3,4}
\end{figure}

Under the RG procedure, the effective Hamiltonian
(\ref{su4vs2::eq:Mixing1}) transforms to the general form
\begin{multline}
H_\eff = J_1\bfs\cdot\bfS + \left[J_3 (t^zT^z) +
4J_2(\bfs\cdot\bfS)(t^zT^z)\right]
\\\mbox{}\quad
+ \left[J_4(t^xT^x+t^yT^y) +
  4J_6(\bfs\cdot\bfS)(t^xT^x+t^yT^y)\right]
\\\mbox{}\quad
+ \left[J_5(t^xT^x-t^yT^y) +
  4J_7(\bfs\cdot\bfS)(t^xT^x-t^yT^y)\right]
\\\mbox{}\quad
+ \left[J_9 + 4J_8(\bfs\cdot\bfS)(t^x+T^x)\right] \,,
\end{multline}
where the exchange coupling constants are initially given by
\begin{multline}
J_1 = J_0 + J_X\,,\quad
J_2 = J_3 = J_0 - J_x\,,\\\mbox{}%
J_4 = J_6 = J_0\,,\quad
J_5 = J_7 = J_X\,,\\\mbox{}%
J_9 = J_8 = \sqrt{J_0J_X} \,.
\end{multline}
Under the RG transformation, they scale as
\begin{subequations}
\label{su4vs2::eq:rg:3}
\begin{align}
\frac{dJ_1}{\rho_0d\ln D} & = -J_1^2 - J_2^2 - 2J_6^2 - 2J_7^2 - 8J_8^2\,, \\
\frac{dJ_2}{\rho_0d\ln D} & = -2J_1J_2 - 2J_4J_6 + 2J_5J_7 \,, \\
\frac{dJ_3}{\rho_0d\ln D} & = -J_4^2 + J_5^2 - 3J_6^2 + 3J_7^2\,, \\
\frac{dJ_4}{\rho_0d\ln D} & = -J_3J_4 - 3J_2J_6\,, \\
\frac{dJ_5}{\rho_0d\ln D} & = J_3J_5 + 3J_2J_7\,, \\
\frac{dJ_6}{\rho_0d\ln D} & = -J_2J_4 - 2J_1J_6 - J_3J_6 - 4J_8^2\,, \\
\frac{dJ_7}{\rho_0d\ln D} & = J_2J_5 - 2J_1J_7 + J_3J_7 - 4J_8^2\,, \\
\frac{dJ_8}{\rho_0d\ln D} & = -2J_1J_8 - 2J_6J_8 - 2J_7J_8 \,,
\end{align}
\end{subequations}
and
\begin{equation}
\label{su4vs2::eq:dJ9} \frac{dJ_9}{d\ln D} = 0 \,.
\end{equation}
As before [see Eq.~(\ref{su4vs2::eq:rg:2})], it is convenient to
work with the reduced coupling constants $j_i\equiv{}J_i/J_1$.  In
terms of these reduced constants, the RG equations read
\begin{subequations}
\label{su4vs2::eq:rg:4}
\begin{align}
\frac{dj_2}{dx} & = -j_2 +
\frac{2j_2 + 2j_4j_6 - 2j_5j_7}{1 + j_2^2 + 2j_6^2 + 2j_7^2 + 8j_8^2} \,, \\
\frac{dj_3}{dx} & = -j_3 +
\frac{j_4^2-j_5^2+3j_6^2-3j_7^2}{1 + j_2^2 + 2j_6^2 + 2j_7^2 + 8j_8^2}\,, \\
\frac{dj_4}{dx} & = -j_4 +
\frac{j_3j_4 + 3j_2j_6}{1 + j_2^2 + 2j_6^2 + 2j_7^2 + 8j_8^2}\,, \\
\frac{dj_5}{dx} & = -j_5 -
\frac{j_3j_5 + 3j_2j_7}{1 + j_2^2 + 2j_6^2 + 2j_7^2 + 8j_8^2}\,, \\
\frac{dj_6}{dx} & = -j_6 +
\frac{j_2j_4+2j_6+j_3j_6+4j_8^2}{1 + j_2^2 + 2j_6^2 + 2j_7^2 + 8j_8^2}\,, \\
\frac{dj_7}{dx} & = -j_7 -
\frac{j_2j_5-2j_7+j_3j_7-4j_8^2}{1 + j_2^2 + 2j_6^2 + 2j_7^2 + 8j_8^2}\,, \\
\frac{dj_8}{dx} & = -j_8 + \frac{2j_8 + 2j_6j_8 + 2j_7j_8}{1 +
j_2^2 + 2j_6^2 + 2j_7^2 + 8j_8^2}
\end{align}
\end{subequations}
together with
\begin{equation}
\label{su4vs2::eq:rg:5}
\frac{1}{(\rho_0J_1)^2}\frac{d(\rho_0J_1)}{d\ln D} = -(1 + j_2^2 +
2j_6^2 + 2j_7^2 + 8j_8^2) \,.
\end{equation}

\begin{figure}
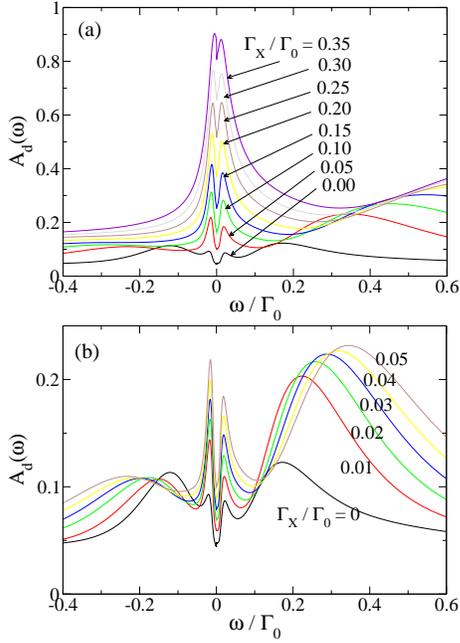

\centering
\includegraphics*[width=6cm]{fig-08a}
\includegraphics*[width=6cm]{fig-08b}
\caption{NRG results for the effects of the finite mixing of the orbital
  quantum number $m$.  The parameter values are: $\eps_{0}=-0.8D$,
  $\Gamma_0=0.08D$, $U_{mm'}=8D$, $\Delta_\orb=32T_K^{SU(4)}$, and
  $\Delta_Z=2T_K^{SU(4)}$.}
\label{su4vs2::fig:nrg-mixing}
\end{figure}

The RG equations (\ref{su4vs2::eq:rg:4}) again have two fixed
points, one associated with the SU(2) Kondo effect and the other
with the SU(4) Kondo effect; see
Fig.~\ref{su4vs2::fig:rgflow-3,4}.  The RG flow diagram in
Fig.~\ref{su4vs2::fig:rgflow-3,4} is reminiscent of that in
Fig.~\ref{su4vs2::fig:rgflow-1,2}.  Both fixed points are stable.
However, since the radius of convergence of the SU(4) Kondo fixed
point is finite, the SU(4) Kondo fixed point cannot be reachable
even for arbitrarily small mixing $V_X$ [see Fig.~\ref{su4vs2::fig:rgflow-3,4}(b)].
However, in the region of interest, the physics is essentially governed by the
SU(4) Kondo fixed point for sufficiently small $V_X$
[see Fig.~\ref{su4vs2::fig:rgflow-3,4}(a)].
Therefore, the SU(4) Kondo physics are in principle unstable against both
the orbital quantum number anisotropy $1-\Gamma_2/\Gamma_1$ and the
orbital mixing $\Gamma_X$.  For sufficiently small values of those,
however, the SU(4) Kondo physics still determines the transport
properties except at extremely low energy scales.  As addressed already,
this suggests that to observe indications of the SU(4)
Kondo physics \emph{exclusively}, one has to investigate the properties
at relatively higher energies (of the order of the Kondo temperature).
This is confirmed and demonstrated in the NRG results summarized in
Fig.~\ref{su4vs2::fig:nrg-mixing}.
We will also see below that there is no way to distinguish the two-level
SU(2) Kondo physics and the SU(4) Kondo physics \emph{experimentally} by
means of linear conductance.

\subsection{Two-Level SU(2) Kondo Effect}

\begin{figure}
\centering
\includegraphics*{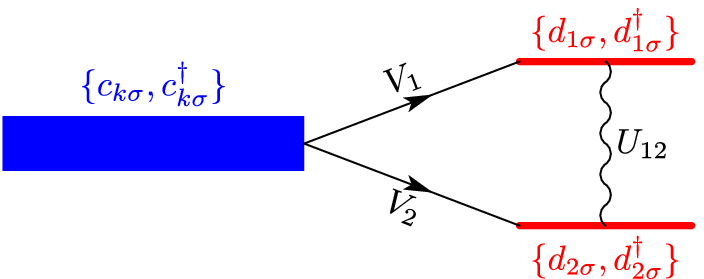}
\caption{(color online) Schematic of the two-level SU(2)-symmetric
  Anderson model.}
\label{su4vs2::fig:model-3}
\end{figure}
As pointed out in the previous subsection, at the maximum mixing
($V_0=V_X$) the physics becomes that of the two-level SU(2) Anderson
model
(see Fig.~\ref{su4vs2::fig:model-3}).
In this case, the only degree of freedom which is conserved during
tunneling is the spin and the total Hamiltonian reads
\begin{multline}
\label{su4vs2::eq:Hsu2-1} H = \sum_{\alpha=L,R}\sum_{k\sigma}
\veps_{\alpha k}a_{\alpha k\sigma}^\dag a_{\alpha k\sigma}
\\\mbox{}
+ \sum_{\alpha km\sigma} V_{m} \left(a_{\alpha k\sigma}^\dag
d_{m\sigma}
  + d_{m\sigma}^\dag a_{\alpha k\sigma}\right)
+ H_D
\end{multline}
or equivalently [see Eq.~(\ref{su4vs2::eq:CT1})]
\begin{multline}
\label{su4vs2::eq:Hsu2-2} H =
\sum_{k\sigma}\veps_kc_{k\sigma}^\dag c_{k\sigma} \\\mbox{}%
+ \sum_{km\sigma}V_m\left(c_{k\sigma}^\dag d_{m\sigma} +
  d_{m\sigma}^\dag c_{k\sigma}\right) + H_D \,.
\end{multline}

As the scaling theory of the Kondo-type Hamiltonian obtained from
the two-level Anderson model has been developed in detail in
Refs.~\onlinecite{Kuramoto98a} and \onlinecite{Eto04a}, we here focus
on the first stage, which highlights the difference between the
two-level SU(2)-symmetric Anderson model and the SU(4)-symmetric
Anderson model. Finally, the physical arguments based on the
perturbative RG theory will be examined quantitatively by means of
the NRG method.

As we integrate out the electronic states in the ranges
$[-D,-(D-\delta{D})]$ and $[D-\delta{D},D]$ of the conduction
band, the dot Hamiltonian (\ref{su4vs2::eq:HD}) evolves as
\begin{multline}
\label{su4vs2::eq:HDR} H_D = \sum_{m\sigma}\eps_{m\sigma}
d_{m\sigma}^\dag d_{m\sigma} - t\sum_\sigma \left(d_{1\sigma}^\dag
d_{2\sigma} + d_{2\sigma}^\dag d_{1\sigma}\right) \\\mbox{}%
+ \sum_{(m,\sigma)\neq (m',\sigma')} U_{mm'}
n_{m\sigma}n_{m'\sigma'}
\end{multline}
with other terms in the total Hamiltonian
(\ref{su4vs2::eq:Hsu2-2}) kept unchanged.  Notice here the
appearance of a new term in $t$, i.e., \emph{a direct transition
between the two orbitals} $m=1$ and $2$.  Scaling of the
parameters $\eps_{m\sigma}$ and $t$ are governed by the RG
equations
\begin{equation}
\label{su4vs2::eq:RG1a} \frac{d\eps_{m\sigma}}{d\ln D} =
-\frac{2}{\pi}\Gamma_m \,
\end{equation}
and
\begin{equation}
\label{su4vs2::eq:RG1b} \frac{dt}{d\ln D} = -
\frac{2}{\pi}\sqrt{\Gamma_1\Gamma_2} \,,
\end{equation}
respectively.

\begin{figure}
\centering
\includegraphics*[width=8cm]{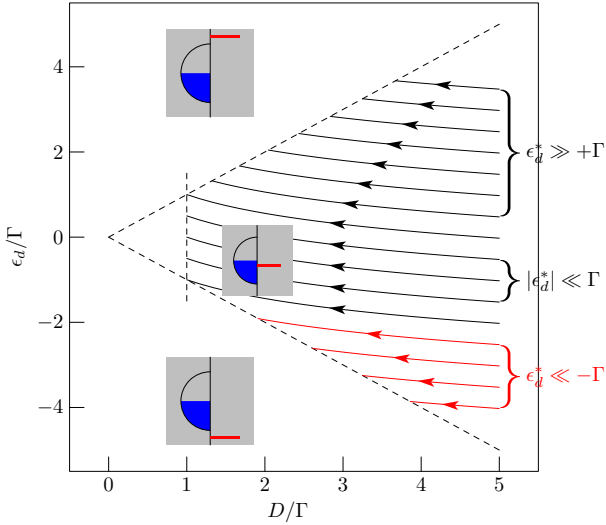}
\caption{(color online) Scaling of the single-particle energy level
  $\eps_d$, to be compared with $\eps_{m\sigma}$ in
  Eq.~(\ref{su4vs2::eq:RG1a}), of the single-level Anderson model.
  $\eps_d^*=\eps_d(D=\Gamma)$ is a scale-invariant quantity.}
\label{su4vs2::fig:scaling-1}
\end{figure}

The RG equation (\ref{su4vs2::eq:RG1a}) for the single-particle
energy levels $\eps_{m\sigma}$ is the same as that in the usual
single-level Anderson model\cite[]{Haldane78a,Haldane78b} (the
corresponding RG flow diagram is shown in
Fig.~\ref{su4vs2::fig:scaling-1}). However, due to the direct
transition $t$ emerging from the RG
equation~(\ref{su4vs2::eq:RG1b}), $\eps_{m\sigma}$ are not
relevant to the Kondo effect [they are not the eigenvalues of
$H_D$ in Eq.~(\ref{su4vs2::eq:HDR})].  To find the relevant energy
level(s) directly involved in the Kondo effect, one may
diagonalize $H_D$ in Eq.~(\ref{su4vs2::eq:HDR}) by means of the
canonical transformation
\begin{equation}
\label{su4vs2::eq:CT2}
\begin{bmatrix}
d_{+,\sigma}\\
d_{-,\sigma}
\end{bmatrix} =
\begin{bmatrix}
\phantom{-}\cos(\theta/2) & \sin(\theta/2) \\
-\sin(\theta/2) & \cos(\theta/2)
\end{bmatrix}
\begin{bmatrix}
d_{1\sigma}\\
d_{2\sigma}
\end{bmatrix} \,,
\end{equation}
where the angle $\phi$ is defined by the relation
\begin{equation}
\tan\theta \equiv \frac{t}{\eps_2-\eps_1} \,.
\end{equation}
The dot Hamiltonian in Eq.~(\ref{su4vs2::eq:HDR}) then
takes the form
\begin{equation}
\label{su4vs2::eq:HDR2} H_D = \sum_{\mu=\pm}\sum_\sigma \eps_\mu
d_{\mu\sigma}^\dag d_{\mu\sigma} +
\sum_{(m,\sigma)\neq(m',\sigma')} U_{mm'}n_{m\sigma}n_{m'\sigma'} \,
\end{equation}
with
\begin{equation}
\label{su4vs2::eq:HDR3} \eps_{\mu=\pm} = \half(\eps_1+\eps_2) \mp
\half\sqrt{(\eps_1-\eps_2)^2 + t^2} \,.
\end{equation}
At the same time, the canonical transformation in
Eq.~(\ref{su4vs2::eq:CT2}) also changes the coupling term in the
total Hamiltonian in Eq.~(\ref{su4vs2::eq:Hsu2-2}) to
\begin{equation}
H_T = \sum_{\mu=\pm}\sum_{k\sigma} V_\mu\left(c_{k\sigma}^\dag
d_{\mu\sigma} +
  d_{\mu\sigma}^\dag c_{k\sigma}\right)
\end{equation}
with $V_\pm$ defined by
\begin{equation}
\label{su4vs2::eq:CT3}
\begin{bmatrix}
V_{+}\\
V_{-}
\end{bmatrix} =
\begin{bmatrix}
\phantom{-}\cos(\theta/2) & \sin(\theta/2) \\
-\sin(\theta/2) & \cos(\theta/2)
\end{bmatrix}
\begin{bmatrix}
V_{1}\\
V_{2}
\end{bmatrix} \,.
\end{equation}
Accordingly, the tunneling rates
\begin{math}
\Gamma_\pm \equiv \pi\rho_0|V_\pm|^2
\end{math}
of the effective orbital levels $\eps_{\pm,\sigma}$ are given by
\begin{equation}
\label{su4vs2::eq:Gamma2} \Gamma_\pm = \half(\Gamma_1+\Gamma_2)
\pm \sqrt{\Gamma_1\Gamma_2}\sin\theta +
\half(\Gamma_1-\Gamma_2)\cos\theta \,.
\end{equation}

\begin{figure}
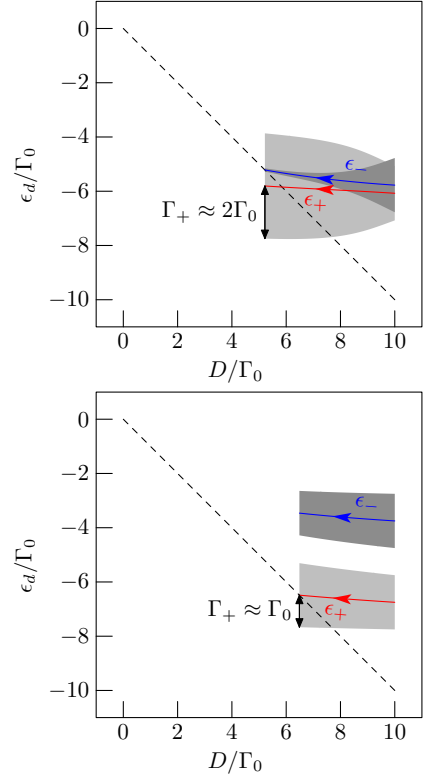

\centering
\includegraphics*[width=55mm]{fig-11a}
\includegraphics*[width=55mm]{fig-11b}
\caption{(color online) Scaling of the two-level SU(2)-symmetric
  Anderson model. The arrowed lines indicate RG flows of the
  effective single-particle energy levels $\eps_\pm$ [see
  Eq.~(\ref{su4vs2::eq:HDR3})] and the widths of the shadowed regions
  around $\eps_\pm$ the RG flow of $\Gamma_\pm$ [see
  Eq.~(\ref{su4vs2::eq:Gamma2})].  The mean of $\Gamma_1$ and $\Gamma_2$
  is denoted by $\Gamma_0$.}
\label{su4vs2::fig:scaling-2,3}
\end{figure}

Figure~\ref{su4vs2::fig:scaling-2,3} shows the scaling of
$\eps_\pm$ (arrowed lines) and $\Gamma_\pm$ (widths of the
shadowed regions around $\eps_\pm$), governed by
Eqs.~(\ref{su4vs2::eq:RG1a}), (\ref{su4vs2::eq:RG1b}),
(\ref{su4vs2::eq:HDR3}), and (\ref{su4vs2::eq:Gamma2}).  Note that
the effective single-particle energy levels $\eps_\pm$ always
repel each other,\cite{Boese02a} and the hybridization
$\Gamma_+$ ($\Gamma_-$) of the lower (upper) level $\eps_+$
($\eps_-$) always increases (decreases).  Essential in this
scaling property of the two-level Anderson model is the direct
transition $t$ between the orbitals $m=1$ and $2$, mediated by the
conduction band.

The scaling of $\eps_\pm$ and $\Gamma_\pm$ stops when the lower
level $\eps_+$ becomes comparable with $D$ ($\eps_+\simeq D$); see
Fig.~\ref{su4vs2::fig:scaling-2,3}.  Then the charge fluctuations
are highly suppressed and the occupation of the lower level becomes
close to unity ($\avg{n_+}\approx 1$).  Therefore, only the lower
level $\eps_+$ gets involved in the Kondo physics, and hence the
resulting Kondo effect is identical to the usual SU(2) Kondo
effect. To be more specific, let us consider the two limiting
cases, $|\eps_1-\eps_2|\gg\Gamma_0$ and
$|\eps_1-\eps_2|\ll\Gamma_0$, assuming
\begin{equation}
|\Gamma_1-\Gamma_2| \ll \Gamma_0\equiv (\Gamma_1+\Gamma_2)/2 \,.
\end{equation}
Since $t\sim\Gamma_0$, one has
\begin{equation}
\theta \approx
\begin{cases}
\pi/2\,, & |\eps_1-\eps_2| \ll \Gamma_0 \,;\\
0 \,, & |\eps_1-\eps_2| \gg \Gamma_0 \,,
\end{cases}
\end{equation}
or equivalently,
\begin{equation}
\Gamma_+ \approx
\begin{cases}
2\Gamma_0 \,, & |\eps_1-\eps_2| \ll \Gamma_0 \,;\\
\Gamma_0 \,, & |\eps_1-\eps_2| \gg \Gamma_0 \,.
\end{cases}
\end{equation}
This implies that when the two orbital levels are nearly
degenerate ($|\eps_1-\eps_2|\ll\Gamma_0$), the Kondo
temperature\cite[]{Haldane78a,Haldane78b} is enhanced
exponentially:
\begin{equation}
\label{su4vs2::eq:TKSU2} T_K\simeq \half\sqrt{2\Gamma_0D}
\exp\left[+\frac{\pi\eps_0}{2\Gamma_0}\right] \,,
\end{equation}
[with $\eps_0\equiv(\eps_1+\eps_2)/2$] compared with the
single-level case (i.e., $|\eps_1-\eps_2|\gg\Gamma_0$)
\begin{equation}
\label{su4vs2::eq:TK0} T_K^0 \simeq \half\sqrt{\Gamma_0D}
\exp\left[+\frac{\pi\eps_0}{\Gamma_0}\right] \,.
\end{equation}

In the limit of nearly degenerate levels
($|\eps_1-\eps_2|\ll\Gamma_0$), the upper level $\eps_-$ is
located at distance smaller than $\Gamma_+$ from the lower level
$\eps_+$ [$(\eps_--\eps_+)\lesssim\Gamma_+$; see the upper
panel in Fig.~\ref{su4vs2::fig:scaling-2,3}] and the transition from
$\eps_+$ to $\eps_-$ is allowed in general.  Indeed, this effect
can be taken into account by a proper SW transformation including
both levels and scaling of the resulting Kondo-like
Hamiltonian,\cite[]{Kuramoto98a,Eto04a} and gives rise to a bump
structure at $\omega=\Delta_\eff$ above the Fermi energy $E_F$ of
the leads, with $\Delta_\eff$ given by \cite[]{Boese02a} (with
$\eps_{1\sigma}=\eps_{2\sigma}$ initially)
\begin{equation}
\label{su4vs2::eq:DeltaEff} \Delta_\eff \sim
\frac{2\Gamma_0}{\pi}\ln\frac{D}{\Gamma_0} \,
\end{equation}
in the single-particle excitation spectrum $A_d(\omega)$ in
Fig.~\ref{su4vs2::fig:nrg-1}; see below.

\begin{figure}
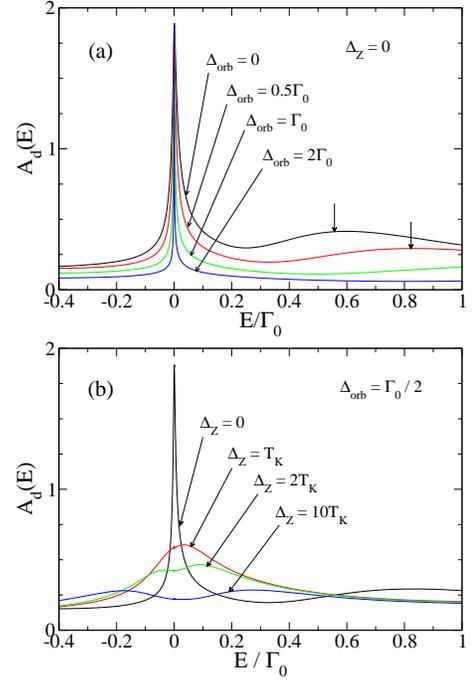

\centering
\includegraphics*[width=6cm]{fig-12a}
\includegraphics*[width=6cm]{fig-12b}
\caption{(color online) Total single-particle excitation spectrum
  $A_d(\omega)$ with (a) only the orbital degeneracy lifted
  ($\Delta_\orb\neq 0$, $\Delta_Z=0$) and (b) both the orbital and spin
  degeneracies lifted ($\Delta_\orb,\Delta_Z\neq0$).  The short vertical
  arrows indicate the transition from $\eps_+$ to $\eps_-$, whose
  excitation energy is given by $\Delta_\eff$ [see
  Eq.~(\ref{su4vs2::eq:DeltaEff})].  The parameter values are: $\eps_{0}=-0.8D$,
  $\Gamma_0=0.1D$, and $U_{mm'}=8D$.}
\label{su4vs2::fig:nrg-1}
\end{figure}

Again, all the interpretations made above on the basis of the
perturbative RG are confirmed with the NRG method.
Figure~\ref{su4vs2::fig:nrg-1} shows the total spectral density
$A_d(E)$.  One can see that as $\Delta_\orb$ increases with
$\Delta_Z=0$, the Kondo peak gets sharper, i.e., the enhancement
of the Kondo temperature $T_K$ in Eq.~(\ref{su4vs2::eq:TKSU2})
diminishes for $\Delta_\orb\geq\Gamma_0$; see
Fig.~\ref{su4vs2::fig:nrg-1}(a). Notice that the bump above the
Fermi energy originates from the excitation via the transition
from the lower level $\eps_+$ to the higher one $\eps_-$, and
is thus located at $E=\Delta_\eff$ [see Eq.~(\ref{su4vs2::eq:DeltaEff}].
When we allow $\Delta_Z$ finite as well,
the Kondo peak then splits into two because of the
Zeeman splitting.\cite{Meir93a,Wingreen94a}

\section{Slave-Boson treatment}
\label{sec:SBT}
In order to confirm our previous results and obtain analytical expressions for intermediate mixing,
we also use slave boson techniques. In particular, the SBMF approach,
which provides a good approximation in
the strong coupling limit $T\ll T_K$, allows us to obtain
analytical expressions for the Kondo temperature and the Kondo
peak position for arbitrary mixing. Our SBMFT results are
complemented with the NCA, which takes
into account both thermal and charge fluctuations in a
self-consistent manner.

At equilibrium it is convenient to change into a representation in
terms of the symmetric (even) and antisymmetric (odd) combinations
of the localized and delocalized orbital channels.~\cite{izu}
Thus the even-odd transformation consists of
$a_{k,1(2),\sigma}=(c_{k_{e}\sigma}\pm i c_{k_{o}\sigma})\sqrt{2}$
and $d_{1(2)\sigma}=(d_{e\sigma}\pm i d_{o\sigma})/\sqrt{2}$.
In this basis the Hamiltonian in Eq.~(\ref{su4vs2::eq:H2}) reads
\begin{multline}
\label{even-odd}
H=\sum_{\sigma,\nu=e,o}\epsilon_{k_{\nu}}c_{k_\nu,\sigma}^\dag
c_{k_\nu,\sigma}+ \sum_{\sigma,\nu=e,o}\veps_{\nu\sigma}
d_{\nu,\sigma}^\dag d_{\nu,\sigma} \\
+ \sum_{\nu=e,o} U n_{\nu\downarrow}
n_{\nu\uparrow} + U n_{e} n_{o}
+V_{e}\sum_{k_e,\sigma}\left(
 c_{k_e,\sigma}^\dag d_{e\sigma}+h.c \right)
\\+ V_{o} \sum_{k_o,\sigma}
\left( c_{k_o,\sigma}^\dag d_{o\sigma} + h.c.\right)\,,
\end{multline}
where, again we have taken $V_0=V_{1,1}=V_{2,2}$,
$V_X=V_{1,2}=V_{2,1}$, $U_{m,m'}=U$, and
$\veps_{\nu,\sigma}=\veps_{0,\sigma}$. The occupation per channel
and spin is given by $n_{\nu\sigma}= d_{\nu\sigma}^\dag
d_{\nu\sigma}$ and the total occupation per channel is
$n_{\nu}=\sum_{\sigma}n_{\nu\sigma}$. 
The tunneling amplitudes for each channel are given by $V_e\equiv V_0+V_X$ and
$V_o\equiv V_0-V_X$. In order to normalize the total tunneling
rate, we take, for the diagonal and off-diagonal tunneling
amplitudes, $V_0=V\cos\phi$ and $V_X=V\sin\phi$, namely
$V_e=V(\cos\phi+\sin\phi)$ and $V_o=V(\cos\phi-\sin\phi)$. Notice
that one needs $\phi\in[0,\pi/4]$ in order to have always $V_{o}$ positively
defined. There exist two very different situations, namely (i) $\phi=0$,
where there are only tunneling processes that conserve
the orbital index, and (ii) $\phi=\pi/4$, where the mixing and
no mixing tunneling amplitudes are the same.

Now we write the physical fermionic operator as a combination of a
pseudofermion operator and a boson operator as follows: $d_{\nu\sigma}=b^\dagger
f_{\nu\sigma}$, where the pseudofermion operator $f_{\nu,\sigma}$
annihilates one ``occupied state'' in the $\nu$th localized orbital
and the boson operator $b^\dagger$ creates an ``empty state''.
Quite generally, the intra-/inter-Coulomb interaction is very large
and we can safely take the limit $U\rightarrow \infty$.
This fact enforces the constraint
\begin{math}
\sum_{\nu\sigma}f_{\nu\sigma}^{\dag}f_{\nu\sigma} + b^\dag b=1
\end{math}
that prevents the accommodation of two electrons at the same time
in either the same orbital or different orbitals. This constraint
is treated with a Lagrange multiplier.
\begin{widetext}
\begin{multline}
\label{hamiltonian1}
H_\mathrm{SB}
= \sum_{\sigma,\nu=e,o}\epsilon_{k_{\nu}}c_{k_\nu,\sigma}^\dag
c_{k_\nu,\sigma}+ \sum_{\sigma,\nu=e,o }\veps_{0,\sigma}
f_{\nu,\sigma}^\dag f_{\nu,\sigma}
+ \frac{\overline{V}_\nu}{\sqrt{N}}\sum_{k,\sigma, \nu=e,o}
\, \left(c_{k_{\nu},\sigma}^\dag b^\dag f_{\nu,\sigma} + h.c.\right)
\\
\mbox{}%
+  \lambda \left(\sum_{\nu,\sigma} f_{\nu,\sigma}^\dag f_{\nu,\sigma}
  + b^\dag b - 1\right) \,.
\end{multline}
\end{widetext}
Notice that we have rescaled the tunneling amplitudes
$V_{e(o)} \to \sqrt{N}\overline{V}_{e(o)}$
according to the spirit of a $1/N$-expansion
(where $N$ is the total degeneracy of the localized orbital).

Our next task is to solve this Hamiltonian, which is rather
complicated due to the presence of the three operators in the
tunneling part and the constrain. In order to do this we employ
two approaches that describe two different physical regimes. The
first one is the SBMFT approach which describes properly the
low-temperature strong coupling regime. This SBMFT provides a good
approximation in the deep Kondo limit, namely in case that only spin
fluctuations are taken into account. The NCA, on the other hand,
takes into account both thermal and charge fluctuations in a
self-consistent manner. It is well known that the NCA fails in
describing the low-energy strong coupling regime. Nevertheless,
the NCA has proven to give reliable results at temperatures down
to a fraction of $T_K$.

\subsection{Slave-boson Mean-Field Theory}\label{sbmft}

We begin with the discussion of the mean-field approximation of
Eq.~(\ref{hamiltonian1}). The merit of this approach is its
simplicity while capturing the main physics in the pure Kondo
regime. It has been successfully applied to investigate the
out-of-equilibrium Kondo effect~\cite{eto01,dong02,ros03,avish03}
and double quantum dots,~\cite{geo99,ram00,ros02,ram03,ros05} just
to mention a few.  This approach corresponds to taking the lowest
order $\mathscr {O}(1)$ in the $1/N$ expansion, where the boson
operator $b(t)$ is replaced by its classical (nonfluctuating)
average, i.e.,
\begin{math}
b(t)/\sqrt{N} \rightarrow \langle b \rangle/\sqrt{N} \equiv
\tilde{b},
\end{math}
thereby neglecting charge fluctuations. In the limit
$N\rightarrow\infty$, this approximation becomes exact. The
corresponding mean-field Hamiltonian is given by
\begin{multline}\label{sbmfHa}
H_\mathrm{MF}= \sum_{\sigma,\nu=e,o}\epsilon_{k_{\nu}}c_{k_\nu,\sigma}^\dag
c_{k_\nu,\sigma}
+ \sum_{\nu,\sigma}\tilde\veps_{0,\sigma}
f_{\nu,\sigma}^\dag f_{\nu,\sigma}
\\\mbox{}%
+ \sum_{\nu,k,\sigma}
\left(\tilde{V}_\nu\,
  c_{\nu,k,\sigma}^\dag f_{\nu,\sigma} + h.c.\right)
\\\mbox{}%
+ \lambda\left(\sum_{\nu,\sigma} f_{\nu,\sigma}^\dag f_{\nu,\sigma}+N|\tilde{b}|^2-1\right) \,,
\end{multline}
where $\tilde{V}_{\nu}=\overline{V}_{\nu}\tilde{b}=V_{\nu}\langle
b \rangle$ and $\tilde{\veps}_{0,\sigma}=\veps_{0,\sigma}+\lambda$
are the renormalized tunneling amplitude and the renormalized
orbital level, respectively. The two mean-field parameters
$\tilde{b}$ and $\lambda$ are to be determined through the
mean-field equations, which are the constraint
\begin{equation}
\label{kondo-su4::eq:premean1} \sum_{\nu,\sigma} \langle
f^\dagger_{\nu,\sigma}(t)f_{\nu,\sigma}(t)\rangle + N|\tilde{b}|^2
= 1
\end{equation}
and the equation of motion for the boson field
\begin{equation}
\label{kondo-su4::eq:premean2}
\sum_{\nu,k,\sigma}\tilde{V}_{\nu}
\left\langle c_{k_{\nu},\sigma}^\dagger(t)f_{\nu,\sigma}(t)\right\rangle
+ \lambda N|\tilde{b}|^2 = 0 \,.
\end{equation}
The Green function for the $\nu \,(= e, o$) localized orbital and
the corresponding lesser lead-orbital Green function are given by
$G^<_{\nu,\sigma}(t-t')=-i \langle f^{\dagger}_{\nu,\sigma}(t')
f_{\nu,\sigma}(t)\rangle$ and
$G^<_{\nu,\sigma,k_{\nu},\sigma}(t-t')=-i\langle
c^\dagger_{k_{\nu},\sigma}(t') f_{\nu,\sigma}(t)\rangle$,
respectively. Expressing the mean-field equations in terms of
these nonequilibrium Green functions, we obtain
Eqs.~(\ref{kondo-su4::eq:premean1}) and
(\ref{kondo-su4::eq:premean2}) in the form
\begin{subequations}
\begin{align}
&\sum_{\sigma}G^{<}_{\nu,\sigma}(t,t) + N|\tilde{b}|^2 = 1\,,
\\
&\sum_{\nu,k,\sigma}\tilde{V}_{\nu}
G^{<}_{\nu,\sigma;k_{\nu},\sigma}(t,t)
+ \lambda N|\tilde{b}|^2 = 0 \,.
\end{align}
\end{subequations}

In order to solve the set of mean-field equations, we proceed as
follows: First, we employ analytic continuation rules to the
equation of motion for the time-ordered Green functions
$G^t_{\nu,\sigma}(t-t')=-i \langle T_C
\{f^{\dagger}_{\nu\sigma}(t')f_{\nu\sigma}(t)\}\rangle$ and
$G^t_{\nu,\sigma;k_{ \nu},\sigma}(t-t')= -i\langle
T_C\{c^\dagger_{k_{\nu},\sigma}(t') f_{\nu,\sigma}(t)\}\rangle$,
where $T_C$ denotes the time-ordering operator along a complex
time contour.~\cite{lan76} Second, we use the equation of motion
technique to relate the lead-orbital Green function with the
orbital Green function. Finally, we rewrite the mean-field
equations in the frequency domain (taking
$\veps_{0,\sigma}=\veps_{0}$):
\begin{subequations}
\label{mean1}
\begin{align}
&|\tilde{b}^2|- \frac{1}{N}\sum_{\nu,\sigma}\int \frac{d\epsilon}{2\pi i}\, G_{\nu\sigma}^<(\epsilon)
= \frac{1}{N}\,, \\
&\lambda|\tilde{b}|^2+ \frac{1}{N}\sum_{\nu,\sigma}\int \frac{d\epsilon}{2\pi i}\,G_{\nu\sigma}^<(\epsilon)
(\veps_{0}-\tilde{\veps}_0)
=0\,.
\end{align}
\end{subequations}
The integrals in Eq.~(\ref{mean1}) can be carried out analytically
by introducing a Lorentzian cutoff $\rho(\eps)=D(\eps^2+D^2)^{-1}$
for the DOS in the leads and the lesser orbital Green function
$G_{\nu}^<(\epsilon)=2i\tilde{\Gamma}_\nu
f(\epsilon)/[(\eps-\tilde{\veps}_0)^2+\tilde{\Gamma}_\nu^2]$ with
$\tilde\Gamma_{\nu}=\overline{\Gamma}_{\nu} |\tilde{b}|^2$ and the
Fermi distribution function $f(\eps)$:
\begin{subequations}
\label{mean2}
\begin{align}
\label{mean2(a)}
&\frac{2}{\pi N}\mathrm{Im} \left [\ln
  \left(\frac{\tilde\veps_0+i\tilde\Gamma_{e}}{D}\right)\right] \nonumber
  +\frac{2}{\pi N}\mathrm{Im} \left [\ln
  \left(\frac{\tilde\veps_0+i\tilde\Gamma_{o}}{D}\right)\right]
\\ &=\frac{1}{N}-|\tilde{b}|^2\,,
\\ \nonumber
\label{mean2(b)}
&\frac{2\tilde\Gamma_{e}}{\pi N}
 {\rm  Re} \left [\ln
  \left(\frac{\tilde\veps_0+i\tilde\Gamma_{e}}{D}\right)\right]+ \frac{2\tilde\Gamma_{o}}{\pi N} {\rm Re} \left [\ln
  \left(\frac{\tilde\veps_0+i\tilde\Gamma_{o}}{D}\right)\right]
\\ &=-\lambda|\tilde{b}|^2\,.
\end{align}
\end{subequations}
In the deep Kondo limit, where $N^{-1}-|\tilde{b}|^2\approx
N^{-1}$ and $-\lambda\approx\veps_0$, these equations obtain the
forms:
\begin{subequations}
\begin{align}
 &{\rm Im} \left [\ln
  \left(\frac{\tilde\veps_0+i\tilde\Gamma_{e}}{D}\right)\right] \nonumber
  +{\rm Im} \left [\ln
  \left(\frac{\tilde\veps_0+i\tilde\Gamma_{o}}{D}\right)\right]
\\ &=\frac{\pi}{2}\,,
\\ \nonumber
 &\Gamma_{e}
 {\rm  Re} \left [\ln
  \left(\frac{\tilde\veps_0+i\tilde\Gamma_{e}}{D}\right)\right]+ \Gamma_{o} {\rm Re} \left [\ln
  \left(\frac{\tilde\veps_0+i\tilde\Gamma_{o}}{D}\right)\right]
\\ &=\frac{\pi\veps_0}{2}\,,
\end{align}
\end{subequations}
where $\Gamma_{\nu}=\overline{\Gamma}_{\nu}/N$ is the original
rate for the $\nu \,(= e, o)$ channel. Using the parametrization
$V_e= V(\cos\phi+\sin\phi)$ and $V_o=V(\cos\phi-\sin\phi)$, the
tunneling rates read $\Gamma_{e}=\pi V^2\rho(1+\sin
2\phi)=\Gamma(1+\sin 2\phi)$ and $\Gamma_{o}=\rho=\pi
V^2\rho(1-\sin 2\phi)=\Gamma (1-\sin 2\phi)$. Taking $\sin
2\phi=\beta$ with $\beta\in[0,1]$ (notice that $0 \leq\sin
2\phi\leq 1$ for $\phi\in[0,\pi/4]$), we parametrize the even and
odd rates as $\Gamma_{e}=(1+\beta)\Gamma$ and
$\Gamma_{o}=(1-\beta)\Gamma$, respectively. Accordingly, the case
$\beta\neq 0$ accounts for the process where even and odd channels
are not coupled equally to the lead electrons or equivalently, the
process where \emph{the orbital index is not conserved}. In terms
of the new notation, the mean-field equations can be written in a
compact way:
\begin{multline}
\label{mean3}
{\rm
  ln}\left[\frac{\tilde{\veps}_0+i\tilde{\Gamma}_{e}}{D}\right]+{\rm
  ln}\left[\frac{\tilde{\veps}_0+i\tilde{\Gamma}_{o}}{D}\right]+{\rm
  ln}\left[\frac{\tilde{\veps}_0^2+\tilde{\Gamma}_{e}^2}{\tilde{\veps}_0^2+\tilde{\Gamma}_{o}^2}\right]^{\beta/2}
  \\
= i\frac{\pi}{2}+\frac{\pi\veps_0}{\Gamma_e+\Gamma_o}\,,
\end{multline}
or equivalently,
\begin{multline}
\label{eq3}
[\tilde{\veps}_0+i\tilde{\Gamma}_{e}][\tilde{\veps}_0+i\tilde{\Gamma}_{o}]\left[\frac{\tilde{\veps}_0^2
+\tilde{\Gamma}_{e}^2}{\tilde{\veps}_0^2+\tilde{\Gamma}_{o}^2}\right]^{\beta/2}
= iD^2 e^{\pi\veps_0 (\Gamma_e+\Gamma_o)^{-1}}\,,
\end{multline}
the real and imaginary parts of which read
\begin{subequations}
\begin{align}
\label{real}
&[\tilde{\veps}_0^2-\tilde{\Gamma}_{e}\tilde{\Gamma}_{o}]\left[\frac{\tilde{\veps}_0^2+\tilde{\Gamma}_{e}^2}{\tilde{\veps}_0^2+\tilde{\Gamma}_{o}^2}\right]^{\beta/2}=0\,,
\\
\label{imaginary} &
\tilde{\veps}_0(\tilde{\Gamma}_{e}+\tilde{\Gamma}_{o})\left[\frac{\tilde{\veps}_0^2+\tilde{\Gamma}_{e}^2}{\tilde{\veps}_0^2+\tilde{\Gamma}_{o}^2}\right]^{\beta/2}=D^2
e^{\pi\veps_0 (\Gamma_e+\Gamma_o)^{-1}}\,.
\end{align}
\end{subequations}
It is obvious that Eq.~(\ref{real}) has the solution
$\tilde{\veps}_0=\pm\sqrt{\tilde{\Gamma}_{e}\tilde{\Gamma}_{o}}$,
among which only the positive root satisfies
Eq.~(\ref{imaginary}). Substituting this result in
Eq.~(\ref{imaginary}), we arrive after some algebra at
\begin{equation}
\label{tkondo0} |\tilde{b}|^2=\frac{D}{\sqrt{2}}\frac{1}{N\Gamma}
\frac{(1-\beta)^{\frac{\beta-1}{4}}}{(1+\beta)^{\frac{\beta+1}{4}}}
\, e^{(\pi\veps_0 /2)(\Gamma_e+\Gamma_o)^{-1}}.
\end{equation}
Using the previous result, we may define the Kondo temperature for
each channel as:~\cite{hew93}
\begin{eqnarray}
\label{kondoevenodd}
T_K^{e}&\equiv&\sqrt{\tilde{\veps}_0^2+\tilde{\Gamma}_{e}^2}
=\frac{(1-\beta)^{\frac{\beta-1}{4}}}{(1+\beta)^{\frac{\beta-1}{4}}} D e^{(\pi\veps_0 /2)(\Gamma_e+\Gamma_o)^{-1}}\,,\nonumber\\
T_K^{o}&\equiv&\sqrt{\tilde{\veps}_0^2+\tilde{\Gamma}_{o}^2}
=\frac{(1-\beta)^{\frac{\beta+1}{4}}}{(1+\beta)^{\frac{\beta+1}{4}}}
D e^{(\pi\veps_0 /2)(\Gamma_e+\Gamma_o)^{-1}}\,,
\end{eqnarray}
and obtain the renormalized level position:
\begin{equation}
\label{relevel} \tilde{\veps}_0= \frac{D}{\sqrt{2}} e^{(\pi\veps_0
/2)(\Gamma_e+\Gamma_o)^{-1}}
\frac{(1-\beta)^{\frac{\beta+1}{4}}}{(1+\beta)^{\frac{\beta-1}{4}}}\,.
\end{equation}
Equations~(\ref{kondoevenodd}) and (\ref{relevel}), which are the
main results of this section, give the Kondo temperatures and
level position for arbitrary mixing $\beta$. Note that
$\Gamma_e+\Gamma_o=2\Gamma$ does not depend on $\beta$ and
therefore the Kondo temperature depends on the orbital mixing only
through the prefactor. While $T_K^{e}$ changes very little with
$\beta$, $T_K^{o}$ reduces \emph{down to zero} as
$\beta\rightarrow 1$ (maximum mixing). Similarly,
$\tilde{\veps}_0$ goes from $T_K\equiv
N\Gamma|\tilde{b}|^2(\beta=0)= (D/\sqrt{2})\exp(\pi\veps_0
/4\Gamma )$ to zero, in agreement with the Friedel sum rule. From
the above results, we conclude that the odd orbital becomes
decoupled at maximum mixing, where we are left with SU(2) Kondo
physics arising from spin fluctuations in the even orbital
channel. This SU(4)-to-SU(2) transition as mixing increases is
illustrated in Fig. 2, where the SBMFT parameters are plotted
versus $\beta$.

Now we are in position to calculate transport properties. For this
purpose, it is more convenient to write SBMFT equations in the
matrix form:
\begin{subequations}
\label{mean4}
\begin{align}\label{eq1}
&|\tilde{b}^2|- \frac{1}{N}\int
\frac{d\epsilon}{2\pi i}\, {\rm Tr}\hat{G}^<(\epsilon)
= \frac{1}{N}\,, \\ \label{eq2}
&\lambda|\tilde{b}|^2+ \frac{1}{N}\int
\frac{d\epsilon}{2\pi i}\,{\rm
  Tr}\{\hat{\Sigma}^r\hat{G}^<(\epsilon)+\hat{\Sigma}^<\hat{G}^a(\epsilon)\}
=0\,,
\end{align}
\end{subequations}
where we are back to the original basis and the trace includes
also the sum over spin indices. Here, $\hat{G}^<$ is the lesser
matrix orbital Green function, which is related to the advanced
$\hat{G}^{a}$ and retarded $\hat{G}^{r}$ matrix Green functions
through the expression
\begin{equation}\label{gles}
\hat{G}^<=\hat{G}^a\hat{\Sigma}^< \hat{G}^r
\end{equation}
with $\hat{\Sigma}^<$ being the lesser matrix self-energy. The
explicit expressions for these matrices are
\begin{eqnarray}\label{ga}
\hat{G}^{a}(\eps)&=&\frac{1}{(\eps-\tilde{\veps}_0-iT_K)^2+\beta^2T_K^2}\\
\nonumber &\times&\left(
\begin{array}{cc}
 \eps-\tilde{\veps}_0-iT_K& i\beta T_K
\\
 i\beta T_K& \eps-\tilde{\veps}_0-iT_K
\end{array} \right)\,
\end{eqnarray}
and $\hat{G}^{r}$ given by direct complex conjugation of
$\hat{G}^{a}$).
The lesser matrix self-energy reads
\begin{eqnarray}\label{sdis}
\hat{\Sigma}^<= - 2 i \left[f_{L}(\eps)+f_{R}(\eps)\right]\times
\left(
\begin{array}{cc}
 T_K &  \beta T_K
\\
 \beta T_K & T_K
\end{array} \right)
\end{eqnarray}
whereas in the same way the retarded matrix self-energy is
\begin{eqnarray}\label{sret}
\hat{\Sigma}^r= -i\hat\Gamma= - i \left(
\begin{array}{cc}
 T_K &  \beta T_K
\\
 \beta T_K & T_K
\end{array} \right)\,.
\end{eqnarray}

Inserting Eqs.~(\ref{ga}) and (\ref{sdis}) in Eq.~(\ref{gles}), we
obtain the lesser orbital Green function:
\begin{widetext}
\begin{eqnarray}\label{gles_explicit}
\hat{G}^<= \frac{ -iT_K}{(\eps-\tilde{\veps}_0)^4+2
(1+\beta^2)T_K^2(\eps-\tilde{\veps}_0)^2+(\beta^2-1)^2T_K^4}
\left(
\begin{array}{cc}
 (\eps-\tilde{\veps}_0)^2+T_K
   (1-\beta^2) &  \beta
   \left[(\eps-\tilde{\veps}_0)^2-T_K (1-\beta^2)\right]\\
\beta\left[(\eps-\tilde{\veps}_0)^2-T_K (1-\beta^2)\right]
&(\eps-\tilde{\veps}_0)^2+T_K
   (1-\beta^2)
\end{array} \right).
\end{eqnarray}
\end{widetext}
Using the explicit expressions of the self-energies and the
nonequilibrium Green functions, we write Eq.~(\ref{eq2}) in the
simple form:
\begin{eqnarray}\label{eq2simplified}
\lambda|\tilde{b}|^2+ \frac{1}{N}\int
\frac{d\epsilon}{2\pi i}\,{\rm
  Tr}\,\hat{G}^<(\epsilon)\,(\eps-\tilde{\veps}_0)
=0\,.
\end{eqnarray}
Solving in a self-consistently way Eqs.~(\ref{eq1}) and
(\ref{eq2simplified}) for each dc bias $V_{\rm dc}$, one gets the
behavior of the two renormalized parameters in nonequilibrium
conditions.\cite{ram00}

\begin{figure}
\centering
\includegraphics*[width=8cm]{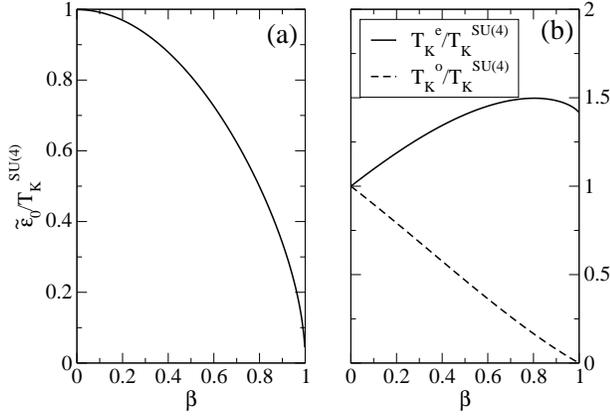}
\caption{(Color online) Transition from SU(4) to SU(2) Kondo
  physics as obtained from SBMFT: As the orbital mixing is increased,
  the SU(4) Kondo effect reduces to the SU(2) spin Kondo effect. This is
  reflected by (a) the position of the Kondo resonance as well as by (b)
  the reduction of the odd Kondo temperature down to zero.
  See the main text.} \label{su4tosu2}
\end{figure}

The electrical current has in appearance the same form as the
conventional Landauer-B\"uttiker current expression for
noninteracting electrons:
\begin{equation}
\label{current}
I=\frac{2e}{\hbar}\int\frac{d\eps}{2\pi}\mathcal{T}(\eps, V_{\rm dc})\left[f_{L}(\eps)-f_{R}(\eps)\right]\,.
\end{equation}
Here caution is needed for a correct interpretation of
Eq.~(\ref{current}), since it contains ``many-body'' effects via
the renormalized parameters that have to be determined for each
$V_{\rm dc}$ in a self-consistent way. As a result, the
transmission $\mathcal{T}(\eps, V_{\rm dc})$ possesses, in
contrast with the noninteracting case, nontrivial behavior with
voltage. The nonlinear conductance is calculated by direct
differentiation of Eq.~(\ref{current}) with respect to the bias
voltage: $\mathcal{G}\equiv dI/dV_{\rm dc}$. In the limit $V_{\rm
dc}\rightarrow 0$ (at equilibrium), the linear conductance
$\mathcal{G}_0$ is given by the well-known expression:
\begin{equation}
\mathcal{G}_0=\frac{2e^2}{h} \mathcal{T}(0)\,,
\end{equation}
where the transmission is
\begin{equation}\label{timplicit}
\mathcal{T}(\eps)={\rm Tr} \{ \hat{G}^a\hat{\Gamma} \hat{G}^r\hat{\Gamma} \}\,.
\end{equation}
Finally, inserting Eqs.~(\ref{ga}) and (\ref{sret}) in
Eq.~(\ref{timplicit}), one arrives at the explicit formula for the
transmission
\begin{widetext}
\begin{eqnarray}
\label{transmission}
\mathcal{T}(\eps)=\frac{2T_K^2\left[(1+\beta^2)(\eps-\tilde{\veps}_0)^2+T_K^2(\beta^2-1)^2\right]}{(\eps-\tilde{\veps}_0)^4+2
(1+\beta^2)T_K^2(\eps-\tilde{\veps}_0)^2+(\beta^2-1)^2T_K^4}\,,
\end{eqnarray}
\end{widetext}
which is the main result of this part. It is remarkable that the
linear conductance $\mathcal{G}_0$ \emph{does not depend on
$\beta$}. In particular, for the SU(4) Kondo model ($\beta=0$),
the transmission takes the simple form:
\begin{equation}
\label{alpha0}
\mathcal{T}(\eps)=\frac{2T_K^2}{(\eps-\tilde{\veps}_0)^2+T_K^2}\,.
\end{equation}
In this case the resonance is pinned at
$\epsilon=\tilde{\veps}_0=T_K$ with the width given by $T_K$; this
leads to $\mathcal{T}(0)=1$ and in consequence
$\mathcal{G}_0=2e^2/h$. For $\beta=1$ corresponding to the
two-level SU(2) Kondo model, Eq.~(\ref{transmission}) reduces to
\begin{equation}
\label{alpha1}
\mathcal{T}(\eps)=\frac{4T_K^2}{(\eps-\tilde{\veps}_0)^2+4T_K^2}\,,
\end{equation}
which leads to the resonance at $\epsilon=\tilde{\veps}_0=0$ and
again $\mathcal{G}_0=2e^2/h$ from $\mathcal{T}(0)=1$. As pointed
out, this fact makes both Kondo effects indistinguishable through
the linear conductance measurement.

\begin{figure}
\centering
\includegraphics*[width=8cm]{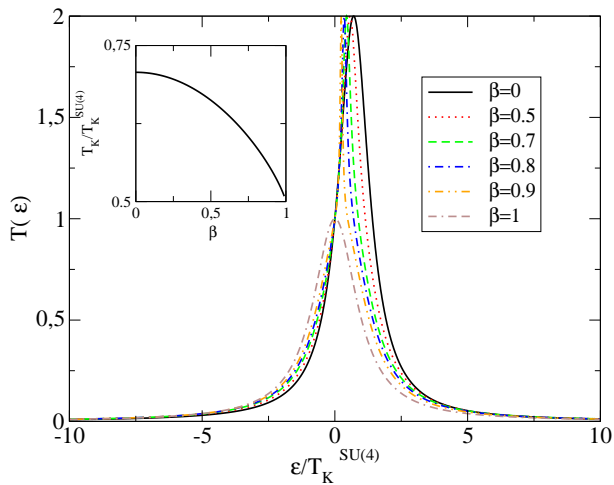}
\caption{(Color online) Equilibrium SBMFT result: Transmission
  $\mathcal{T}(\eps)$ as a function of the frequency for several values
  of $\beta$. The left inset displays the Kondo temperature versus
  $\beta$.}
\label{slave1}
\end{figure}

All these features are clearly observed in Fig.~\ref{slave1},
where the transmission for different amounts of mixing, i.e.,
different values of $\beta$ is plotted. For $\beta=0$ the
transmission peak is located at $T_K$ as expected, whereas for
$\beta=1$ this moves towards $\eps=0$. During the transition from
the SU(4) to the two-level SU(2) Kondo model, the transmission
gets narrower and develops a ``cusp'', signaling competition
between even and odd channels. This is manifested by the following
from of the transmission:
\begin{widetext}
\begin{equation}
\label{interference} \mathcal{T}(\eps)
=\frac{(1+\beta)^2T_K^2}{(\eps-\tilde{\veps}_0)^2+T_K^2(1+\beta)^2}
+\frac{(1-\beta)^2T_K^2}{(\eps-\tilde{\veps}_0)^2+T_K^2(1-\beta)^2}\,.
\end{equation}
\end{widetext}
Note that both channels are resonant at the same energy
$\tilde{\veps}_0$ but have different widths ($\tilde{\Gamma}_{e}$
and $\tilde{\Gamma}_{o}$), which explains the ``cusp'' behavior.
Here we speculate that finite splitting $\delta\veps$ originating
from charge fluctuations \cite{Manhn-SooPRL04} (not included at
the SBMFT level) would give rise to two split resonances for
$\beta\neq 0$, namely,
$\tilde{\veps}_0\rightarrow\tilde{\veps}_0^{\pm}=\delta\veps\pm\sqrt{\tilde{\Gamma}_{e}\tilde{\Gamma}_{o}}$.
This is confirmed in the next section, where we present results
obtained from full NCA calculations including fluctuations.
Eventually, for $\beta=1$ the competition does not exist and the
transmission does not display the cusp.

\subsection{Non-crossing approximation Method}
The SBMFT suffers from two drawbacks: 1) It leads always to local
Fermi liquid behavior and 2) there arises a phase transition
(originating from breakdown of the local gauge symmetry of the
problem) that separates the low-temperature state from the
high-temperature local moment regime. The latter problem may be
corrected by including 1/N corrections around the mean-field
solution. The non-crossing approximation (NCA)
\cite{NCAneq1,NCAneq2,NCAneq3} is the lowest-order
self-consistent, fully conserving, and $\Phi$ derivable theory in
the Baym sense \cite{Baym} which includes such corrections.
Without entering into details of the theory, we just mention that
the boson fields in Eq.~(\ref{hamiltonian1}), treated as averages
in the previous subsection ($b(t)/\sqrt{N} \rightarrow \langle b
\rangle/\sqrt{N} \equiv \tilde{b}$), are now treated as
fluctuating quantum objects. In particular, one has to derive
self-consistent equations of motion for the time-ordered
double-time Green function (with subindexes omitted):
\begin{eqnarray}
iG(t,t')&\equiv&\langle T_c f(t)f^\dagger(t')\rangle\nonumber\,,\\
iB(t,t')&\equiv&\langle T_c b(t)b^\dagger(t')\rangle,
\end{eqnarray}
or in terms of their analytic pieces,
\begin{eqnarray}
iG(t,t')&=&G^{>}(t,t')\theta(t-t')-G^{<}(t,t')\theta(t'-t)\,,\nonumber\\
iB(t,t')&=&B^{>}(t,t')\theta(t-t')+B^{<}(t,t')\theta(t'-t).
\end{eqnarray}

\begin{figure}
\centering
\includegraphics*[width=8cm]{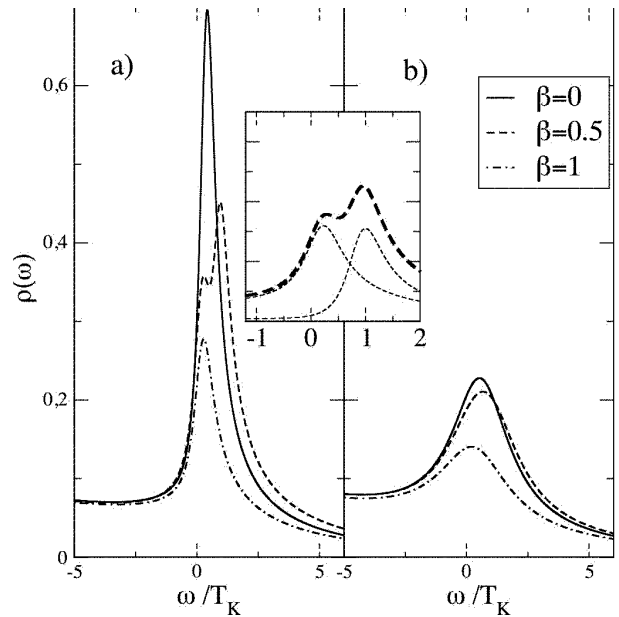}
\caption{(Color online) NCA results: Density of states around
  $\varepsilon=0$ at $T=0.25T_K$ (left) and $T=T_K$ (right) for
  several values of $\beta$. The inset shows a close-up of the
  $\beta=0.5$ curve (thick dashed), together with the individual even and
  odd channel contributions (thin dashed).} \label{nca1}
\end{figure}

A rigorous and well-established way to derive these equations of
motion was introduced,\cite{Kadan} and related to other
non-equilibrium methods such as the Keldysh method.\cite{lan76}
Here, we just present numerical results of the NCA equations for
our problem and refer interested readers to
Refs.~\onlinecite{ram03,NCAneq1,NCAneq2,NCAneq3} for details.

In particular, the DOS is given by
\begin{equation}
\rho(\varepsilon)=-\frac{1}{\pi}\sum_{\nu=e,o,\sigma}
\mathrm{Im}[A^{r}_{\nu\sigma}(\varepsilon)],
\end{equation}
where $A^{r}_{\nu\sigma}(\varepsilon)$ is the Fourier transform of
the retarded Greens function:
\begin{equation}
A^{r}_{\nu\sigma}(t)=
G_{\nu,\sigma}^{r}(t)B^{<}(-t)-G^{<}_{\nu\sigma}(t) B^{a}(-t).
\end{equation}
The DOS for several values of $\beta$ at two different
temperatures is plotted in Fig.~\ref{nca1}. Interestingly, the
cusp behavior of Fig.~\ref{slave1} in the previous subsection
becomes split for the even and odd channels. This is illustrated
in the inset, where the curve corresponding to $\beta=0.5$ is
plotted together with the individual even and odd channel
contributions. As we anticipate, the presence of charge
fluctuations induces splitting of $\veps_0$ due to the different
renormalization arising from different couplings for the even and
odd channels $\Gamma_{e/o}$ [see Eqs.~(\ref{su4vs2::eq:RG1a}) and
(\ref{su4vs2::eq:RG1b})].

\section{Conclusion}
\label{sec:conclusion}

We have considered the single-electron transistor (SET) device
with the CNT QD or VQD as the central island in the Kondo regime.
In particular we have examined the case where the CNT QD or VQD
has a high symmetry so that the orbital quantum numbers are
conserved through the system.  Emphasis has been paid on how
different Kondo physics, the SU(4) Kondo effect or the two-level
SU(2) Kondo effect, emerges depending on the extent to which the
symmetry is broken in realistic situations.  Employed are four
different theoretical approaches: the scaling theory, the NRG
method, the SBMFT, and the NCA method to address both the linear
and non-linear conductance for given external magnetic fields.
Our results show that there is no way to distinguish
experimentally the SU(4) Kondo effect and the two-level SU(2)
effect by means of the linear conductance (which is a low-energy
property) alone.  The SU(4) Kondo physics, which arises with
higher symmetry, can be observed exclusively only by the
non-linear conductance (a higher-energy property) in the presence
of finite magnetic fields, as in the recent
experiment.\cite{Jarillo-Herrero05a,Jarillo-Herrero05b}
The symmetry breaking (either the orbital anisotropy
$1-\Gamma_2/\Gamma_1$ or the orbital mixing $\Gamma_X$) drives the
system from the SU(4) Kondo fixed point to the SU(2) Kondo fixed
point. At finite yet sufficiently small symmetry breaking, the
SU(4) Kondo physics governs the transport in the system at
relatively high energies (of the order of the Kondo temperature)
while the two-level SU(2) Kondo effect takes it over at extremely
low energies.  This gives another reason why the indication of the
SU(4) Kondo physics should be investigated by means of the
non-linear conductance in the presence of external magnetic
fields.

\section*{Acknowledgments}

This work was supported by the MEC of Spain (Grants FIS2005-02796,
MAT2005-07369-C03-03, the Ramon y Cajal Program), CSIC (Proyecto
Intramural Especial I3), the SRC/ERC program of MOST/KOSEF
(R11-2000-071), the Korea Research Foundation Grant
(KRF-2005-070-C00055), the BK21 Program, and the SK Fund.

\appendix
\section{Numerical Renormalization Group}

The Hamiltonian (\ref{su4vs2::eq:H2}) allows both charge
fluctuations and spin fluctuations.  Charge fluctuations
(accompanied by the particle-hole excitations) occurs at high
energies while spin fluctuations prevail at low energies.
Therefore in order to understand the low-energy properties of the
system, it is useful to take the RG approach and to obtain an
effective Hamiltonian allowing only the spin fluctuations.  One
may follow the three-state perturbative RG procedure (scaling
theory): One first renormalizes the Anderson-type Hamiltonian
(\ref{su4vs2::eq:H2}) until the charge fluctuations are completely
suppressed\cite[]{Haldane78a,Haldane78b} (see also
Ref.~\onlinecite{Boese02a}), performs the SW
transformation\cite[]{Schrieffer66a} to obtain a Kondo-type
Hamiltonian where spin fluctuations are described by the spin
operators, and renormalizes further the resulting Kondo-type
Hamiltonian.\cite[]{Anderson70a}  The RG equations for the
coupling constants in the Hamiltonian allow one to identify
physically interesting fixed points and associated scaling
properties.

We follow the standard procedure to implement the NRG
calculations,~\cite{Wilson75a,Krishna-murthy80a,Costi94a} and
evaluate various physical quantities from the recursion relation
($N\geq 0$)
\begin{multline}
\label{sds-kondo::eq:HNRG} \tilH_{N+1} = \sqrt\Lambda\, \tilH_N
\\\mbox{} + \xi_{N+1}\sum_{\mu\sigma} \left(f_{\mu,N,\sigma}^\dag
f_{\mu,N+1,\sigma} + h.c.\right)
\end{multline}
with the initial Hamiltonian
\begin{equation}
\label{sds-kondo::eq:H0} \tilH_0 = \frac{1}{\sqrt\Lambda} \Biggl[
\tilH_D + \sum_{\mu m}\sum_\sigma\tilV_{\mu,m} \left(
f_{\mu,0,\sigma}^\dag d_{m\sigma} + h.c. \right) \Biggr] \,,
\end{equation}
where the fermion operators $f_{\mu,N,\sigma}$ have been
introduced as a result of the logarithmic discretization and the
accompanying canonical transformation, $\Lambda$ is the
logarithmic discretization parameter (taken to be $\Lambda=2$),
\begin{equation}
\xi_N \equiv
\frac{1-\Lambda^{-N}}{\sqrt{[1-\Lambda^{-(2N-1)}][1-\Lambda^{-(2N+1)}]}}
\,,
\end{equation}
and
\begin{equation}
\tilH_D \equiv \zeta \frac{H_D}{D}
\end{equation}
with $\zeta=2(1+1/\Lambda)^{-1}$. The coupling constants
$\tilV_{\mu,m}$ have been defined to be
\begin{gather}
\tilV_{\mu,m} \equiv \sqrt{\frac{2\rho_0|V_{\mu,m}|^2}{\pi D}} \,,
\end{gather}
where $\rho_0$ is the density of states of the leads at the Fermi
energy. 
The Hamiltonian $\tilH_N$ in Eq.~(\ref{sds-kondo::eq:HNRG}) has
been rescaled for numerical accuracy, and the original Hamiltonian
is recovered by
\begin{equation}
\frac{H}{D} = \lim_{N\to\infty}\frac{\tilH_N}{\varS_N}
\end{equation}
with
\begin{math}
\varS_N \equiv \zeta\Lambda^{(N-1)/2}.
\end{math}
At each iteration of the NRG procedure, we calculate the local
spectral density,\cite{Bulla01a} which determines the transport
properties through the dot:
\begin{equation}
\label{su4vs2::eq:2-21} A_{mm'}(\omega) = A_{mm'}^>(\omega) -
A_{mm'}^<(\omega)
\end{equation}
with
\begin{subequations}
\label{su4vs2::eq:2-22}
\begin{align}
\label{su4vs2::eq:2-22a} A_{mm';\sigma}^>(E) & = + \sum_\alpha
\bra{0}d_{m\sigma}\ket{\alpha}
\bra{\alpha}d_{m'\sigma}^\dag\ket{0}
\delta(E - E_\alpha + E_0) \,,\\
\label{su4vs2::eq:2-22b} A_{mm';\sigma}^<(E) & = - \sum_\alpha
\bra{0}d_{m'\sigma}^\dag\ket{\alpha}
\bra{\alpha}d_{m\sigma}\ket{0} \delta(E + E_\alpha - E_0) \,,
\end{align}
\end{subequations}
where $\ket{\alpha}$ represents the many-body state of the system
with energy $E_\alpha$ (with $\alpha = 0$ corresponding to the
ground state).


\end{document}